\documentclass[journal,draftclsnofoot,12pt,onecolumn,english]{IEEEtran}
\usepackage[T1]{fontenc}
\usepackage{amsmath,amssymb}
\usepackage{graphicx, array,booktabs}
\usepackage{microtype}
\usepackage{tikz,pgfplots}
\pgfplotsset{compat=1.18}
\usepackage{epstopdf}
\usetikzlibrary{arrows.meta}
\usepackage{placeins}
\usepackage{xcolor}
\newcommand{\blue}{\color{blue}}

\usepackage{cite}
\usepackage{pstricks}
\usepackage{multirow}

\newcommand{\tabincell}[2]{\begin{tabular}{@{}#1@{}}#2\end{tabular}}
\usepackage[hidelinks]{hyperref}
\hypersetup{pdftitle={Exact Delivery for Cyclic L-Separated Placement in Multi-Access Coded Caching: Converse, Determinant-Based Coding, and Global Bounds},pdfauthor={Anonymous Authors},pdfsubject={Determinant-based linear coding and Schwartz-Zippel analysis}}
\newtheorem{theorem}{Theorem}
\newtheorem{corollary}{Corollary}
\newtheorem{lemma}{Lemma}

\newtheorem{remark}{Remark}

\newcommand{\pos}[1]{\left(#1\right)^{+}}
\newcommand{\cyc}[1]{\left\langle #1\right\rangle_K}

\usepackage{bm}

\makeatletter
\def\@begintheorem#1#2{%
  \@IEEEtmpitemindent\itemindent\relax
  \topsep 0pt
  \rmfamily
  \trivlist
  \item[]\indent
  \textbf{#1\ #2}:\nobreak\hspace{0.5em}%
  \itemindent\@IEEEtmpitemindent\relax
}

\def\@opargbegintheorem#1#2#3{%
  \@IEEEtmpitemindent\itemindent\relax
  \topsep 0pt
  \rmfamily
  \trivlist
  \item[]\indent
  \textbf{#1\ #2}\textit{\ (#3)}:\nobreak\hspace{0.5em}%
  \itemindent\@IEEEtmpitemindent\relax
}
\makeatother
\allowdisplaybreaks

\begin{document}
% Reconstructed front matter and Sections I--III. The body uses only
% determinant coefficients and ordinary finite-field linear algebra.
\title{The Exact Optimal Delivery Scheme for the Cyclically $L$-Separated Placement in Multi-Access
Coded Caching}
\author{
Kai~Wan,~\IEEEmembership{Member,~IEEE,} 
Huimei~Wei, 
~Minquan~Cheng,~\IEEEmembership{Member,~IEEE,}  
and~Giuseppe~Caire,~\IEEEmembership{Fellow,~IEEE}
\thanks{
K.~Wan and H.~Wei are with the School of Electronic Information and Communications,
Huazhong University of Science and Technology, 430074  Wuhan, China (e-mail: \{kai\_wan, Scarlett\_w\}@hust.edu.cn).}
\thanks{M.~Cheng is with the Key Laboratory of Education Blockchain and
Intelligent Technology, Ministry of Education, and Guangxi Key Laboratory of
Multi-Source Information Mining and Security, Guangxi Normal University,
541004 Guilin, China (e-mail: chengqinshi@hotmail.com).}
\thanks{G.~Caire is with the Electrical Engineering and Computer Science Department, Technische Universit\"at Berlin, 10587 Berlin, Germany (e-mail: caire@tu-berlin.de). %The work of G. Caire was supported by the Gottfried Wilhelm Leibniz-Preis 2021 of the German Science Foundation (DFG).
}
}
\maketitle

\begin{abstract} 
This paper investigates the $(K, L, M, N)$ multi-access coded caching (MACC) system introduced by Hachem \textit{et al.} The system consists of a server storing $N$ equal-sized files, $K$ cache nodes, each with a storage capacity of $M$ files, and $K$ users, each accessing $L$ consecutive cache nodes in a cyclic wrap-around topology.  The server transmits coded information to the users via an error-free capacitated broadcast channel, such that each user obtains its own requested file as a combination of the server transmission and the accessible caches’ content. The goal is to minimize the worst-case transmission load over the server-to-users broadcast channel over all possible user demands. 
Our objective is to minimize the worst-case transmission load over all possible user demands. 
First, for the cyclically $L$-separated placement, we derive a new converse bound using the acyclic index coding bound. By carefully counting all valid cache supports, the proposed converse improves upon an existing bound for the same placement.  
We then develop a linear coded caching scheme based on a novel determinant-based interference alignment strategy, which achieves the new converse bound exactly. Exploiting the determinant structure of a random generating matrix (each of whose columns corresponds to one user), this scheme simultaneously aligns interference across all users into low-dimensional subspaces, while preserving the decodability of their desired packets.  
Furthermore, by averaging acyclic index coding bounds over cyclic user permutations, we derive a general converse bound applicable to arbitrary uncoded cache placements. By combining the proposed scheme with memory sharing, we demonstrate that the achievable transmission load is within a factor of two of the global bound under uncoded placement. Finally, theoretical and numerical comparisons demonstrate that the proposed scheme generally achieves a strictly lower load than existing schemes.
\end{abstract}


\section{Introduction}
\label{Section-Intro}

Coded caching, introduced by Maddah-Ali and Niesen \cite{MN}, is an effective technique for reducing the transmission load in content delivery networks. By jointly designing data storage during off-peak hours and coded transmissions during peak hours, it enables a single coded multicast transmission to serve multiple users simultaneously, even when they request different files. It has been widely studied in various network settings, including device-to-device (D2D) networks~\cite{Ji2013}, hierarchical networks~\cite{Karamchandani2016}, wireless networks~\cite{Hachem2015}, multi-access networks~\cite{HKD}, and combination networks~\cite{Ji2015Combination}. 

Among these settings, multi-access coded caching (MACC) was first introduced by Hachem \textit{et al.}~\cite{HKD} to model edge caching in heterogeneous wireless networks. The motivating architecture combines densely deployed access points, such as small-cell base stations or Wi-Fi access points, that provide high data rates over short distances with sparsely deployed macro base stations that provide wider coverage at lower rates per link. By equipping the access points with caches, users can recover their requested files using both cached content retrieved from multiple nearby access points and broadcast transmissions from a macro base station. To reflect users' access to nearby caches, Hachem \textit{et al.}~\cite{HKD} considered a system model in which each user can retrieve cache content from consecutive caches, with cyclic wrap-around for symmetry.

In a $(K, L, M, N)$ cyclical MACC system, a server storing $N$ equal-sized files serves $K$ users without local caches over an error-free shared link. There are $K$ cache nodes, each with capacity equal to the size of $M$ files. User $k$ accesses the $L$ consecutive cache nodes indexed by $k, k+1, \ldots, k+L-1$, with indices interpreted cyclically over $\{1,2,\ldots,K\}$.
A MACC scheme consists of two phases: placement during off-peak hours and delivery during peak hours. During the placement phase, the server places data of total size at most $M$ times the file size in each cache, without knowledge of future user demands. If file bits are stored directly without coding, the placement is called uncoded placement.
During the delivery phase, each user randomly requests one file from the library. Given these demands, the server broadcasts coded messages that enable each user to recover its requested file using the received messages and the contents of its accessible caches. The transmission load $R$ is defined as the maximum total amount of data broadcast by the server over all possible user demands, normalized by the file size. Our objective is to jointly design cache placement and delivery to minimize $R$ for a given cache size $M$.
\subsection{Case \texorpdfstring{$L=1$}{L=1}}
When $L=1$, each user accesses only its corresponding cache, and the model reduces to the classical $(K, M, N)$ coded caching system in~\cite{MN}. Maddah-Ali and Niesen~\cite{MN} derived a cut-set lower bound on the transmission load and proposed the first coded caching scheme, referred to as the MN scheme. For $N\ge K$ and $M/N=t/K$, where $t\in\{0,\ldots,K\}$, the MN scheme achieves a transmission load of $R=(K-t)/(t+1)$. By comparing this achievable load with the cut-set lower bound, they showed that its load is at most $12$ times the information-theoretic optimum.

There are many studies on improving the cut-set lower bound derived in~\cite{MN}, both for general placement~\cite{Ajaykrishnan2015, Sengupta2015, Wang2016Converse, Ghasemi2017, Tian2018, Wang2018Converse, Yu2019Converse} and under the restriction of uncoded placement~\cite{WTP2016, YMA2018, WTP2020, Krishnan2021}. Ajaykrishnan \textit{et al.}~\cite{Ajaykrishnan2015} obtained tighter bounds by jointly analyzing cache contents and transmissions for different demand vectors. For $N=K=3$ and $M=1$, their bound is $3/2$ times the cut-set bound, increasing the lower bound from $2/3$ to $1$ and matching the MN load. Sengupta \textit{et al.}~\cite{Sengupta2015} strengthened the cut-set bound using Han's inequality~\cite[Thm.~17.6.1]{CoverThomas2006}. Wang \textit{et al.}~\cite{Wang2016Converse} derived additional converse bounds by reducing cooperating users to a single user requesting multiple files. Ghasemi and Ramamoorthy~\cite{Ghasemi2017} developed a labeling algorithm on directed trees, yielding a bound that includes the cut-set bound as a special case and is strictly tighter in certain parameter regimes. Tian~\cite{Tian2018} exploited problem symmetry in entropy linear programming to characterize the exact trade-off for two-user systems and obtain further bounds for other small systems. Wang \textit{et al.}~\cite{Wang2018Converse} derived improved bounds using conditional mutual information inequalities and Han's inequality. Yu \textit{et al.}~\cite{Yu2019Converse} strengthened the cut-set argument by exploiting the individual decoding constraints of users on the same side of a cut. Under uncoded placement, Wan \textit{et al.}~\cite{WTP2016, WTP2020} applied the acyclic index coding bound to prove the optimality of the MN scheme for $N\ge K$. Yu \textit{et al.}~\cite{YMA2018} characterized the optimal memory–load trade-off for arbitrary $N$ and $K$ using a delivery scheme that removes redundant transmissions arising from repeated requests and a matching genie-aided converse. Krishnan \textit{et al.}~\cite{Krishnan2021} developed a general data-exchange converse using induction, recovering several known uncoded-placement bounds.

However, the MN scheme requires $\binom{K}{t}$ equal-sized packets per file, a subpacketization that grows exponentially with $K$ for a fixed cache ratio $M/N\in(0,1)$. There are many studies on reducing subpacketization relative to that of the MN scheme under uncoded placement and one-shot delivery, where one-shot means that each required packet is transmitted by the server exactly once.  For instance, Shanmugam \textit{et al.}~\cite{SJTLD} applied coded caching within smaller user groups, and Cheng \textit{et al.}~\cite{CJWY} proposed a generalized grouping scheme. Yan \textit{et al.}~\cite{YCTC} introduced the placement delivery array (PDA) framework to describe uncoded placement and one-shot linear delivery in a single array. They also showed that the MN scheme can be represented by a PDA, referred to as the MN PDA. Cheng \textit{et al.}~\cite{CJYT, CJTY} developed further PDA constructions, and Cheng \textit{et al.}~\cite{CWZW} constructed PDAs using orthogonal arrays. Zhong \textit{et al.}~\cite{ZCW2021} used Hamming distance to construct schemes with linear subpacketization. Aravind \textit{et al.}~\cite{AST2022} developed lifting constructions. Wang \textit{et al.}~\cite{WCWC2022ISIT, WCWC2025} constructed PDAs using Cartesian products and unions of the resulting cache configurations. For graph-based constructions, Yan \textit{et al.}~\cite{YTCC} represented PDAs through strong edge colorings of bipartite graphs, and Michel and Wang~\cite{MW2020} combined strong edge colorings to construct new PDAs. Shangguan \textit{et al.}~\cite{SZG} developed a hypergraph approach, Shanmugam \textit{et al.}~\cite{STD2017} used Ruzsa–Szemerédi graphs to obtain linear subpacketization, and Krishnan~\cite{K2018} proposed constructions based on line graphs of bipartite graphs. Combinatorial designs and finite geometry have also been used to construct coded caching schemes. Tang and Ramamoorthy~\cite{TR2018} constructed resolvable designs from linear block codes, and Agrawal \textit{et al.}~\cite{ASK} developed schemes based on combinatorial designs. Chittoor \textit{et al.}~\cite{CKSB2021} used projective geometry to obtain schemes with subexponential or linear subpacketization, while Cheng \textit{et al.}~\cite{CWWC2026} used combinatorial packings to construct schemes with linear subpacketization.

\subsection{General case \texorpdfstring{$L>1$}{L>1}} 
For $L>1$, existing studies~\cite{HKD, NR, Vijith2023MACC, RK, CWLZC, RK2, SPE, SR_arX2021, MR, KK2023, SR, WCWL, LCWC} have focused on deriving lower bounds on the transmission load and constructing coded caching schemes.

For general placement, Hachem \textit{et al.}~\cite{HKD} derived an information-theoretic lower bound on the transmission load using a sliding-window subset entropy inequality. For $L\mid K$, they proposed a scheme by partitioning the users into $L$ groups and applying the MN scheme within each group. They showed that its transmission load is at most $cL$ times the information-theoretic optimum, where $c$ is independent of the system parameters. Namboodiri and Rajan~\cite{NR} derived tighter information-theoretic lower bounds using cut-set arguments and a sliding-window subset entropy inequality, and characterized the optimal memory–load trade-off for $N=K=3$ and $L=2$. Vijith Kumar \textit{et al.}~\cite{Vijith2023MACC} derived lower bounds and constructed a matching scheme for $L=K-2$ and $M\in[N(K-1)/(K(K-2)),N/(K-2)]$, thereby characterizing the optimal memory–load trade-off in this regime. Under uncoded placement, Reddy and Karamchandani~\cite{RK} applied the acyclic index coding bound to derive a converse for $L\ge K/2$. They also proposed an index-coding-based scheme, referred to as RK1, for arbitrary $L$, and showed that its transmission load is at most twice the optimal load under uncoded placement when $L\ge K/2$. Cheng \textit{et al.}~\cite{CWLZC} derived a converse for a fixed cyclically $L$-separated placement and developed a transformation from PDAs satisfying specified conditions to cyclic MACC schemes while preserving their coded caching gains. Based on the MN PDA, they obtained the CWLZC scheme, which achieves the HKD load expression without requiring $L\mid K$ and is asymptotically optimal under its fixed placement as $K$ grows with $L$ and $KM/N$ fixed. 

Other studies have focused on constructing schemes with lower transmission loads~\cite{RK2, SPE, SR_arX2021}. Reddy and Karamchandani~\cite{RK2} developed the RK2 scheme using structured index coding. RK2 retains the placement of RK1 and achieves a load no greater than that of either RK1 or CWLZC. By comparing its achievable load with the converse in~\cite{RK}, they showed that its load is at most $5/4$ times the optimal load under uncoded placement for $L\ge K/2$ and $M=N/K$. Serbetci \textit{et al.}~\cite{SPE} jointly designed placement and delivery to achieve coded caching gains exceeding $t+1$ for certain parameters, where $t=KM/N$. Sasi and Rajan~\cite{SR_arX2021} proposed the SR1 scheme for coprime $t$ and $K$, improving upon the load of RK1 for certain parameters and attaining the optimal load when $K-tL=1$. Table~\ref{tab_schemes} summarizes the transmission loads and parameter constraints of the schemes considered in our performance comparisons. Other studies have focused on reducing subpacketization, at the cost of a higher transmission load, using index coding~\cite{MR, KK2023}, PDAs~\cite{SR,WCWL}, and extensions of the combinatorial packings in~\cite{CWWC2026} to cyclical MACC systems~\cite{LCWC}.

Despite numerous studies and the resulting improvements, a gap remains between the known converse bounds and achievable loads for cyclical MACC systems with general parameters~\cite{RK, RK2, CWLZC}. Deriving tight converse bounds and constructing delivery schemes that attain them, either exactly or asymptotically, therefore remain important problems. A key difficulty is that users access overlapping sets of caches, so the side information available to each user depends jointly on the cache placement and the cyclic access pattern. Even when the placement is fixed, a tight converse must account for this side-information structure, while a matching delivery scheme must exploit it to satisfy all users' decoding requirements simultaneously.

Beyond cyclic access, MACC has been studied under other access topologies~\cite{Katyal2021CRD, Das2022CRD, Zhang2024TwoDim, Brunero2023Combinatorial, Cheng2025Topology}. Multi-antenna extensions have also been investigated~\cite{Peter2023MultiAntenna, Cheng2024MAMISO, Peter2025MultiAntenna}.

\begin{table*}[htbp!]
        \centering
        \renewcommand{\arraystretch}{1.7}
        \setlength{\tabcolsep}{3pt}
        \caption{Known multi-access coded caching schemes with uncoded placement. Here $K$ is the number of users, the number of cache nodes equals $K$, $L$ is the access degree, and $t/K$ is the normalized cache-node memory size. The notation $ \|\mathbf b\|_\infty$ denotes the infinity norm of the vector $\mathbf b$.  Unless otherwise specified in the constraints column,  $t\in \{1,\ldots,\lfloor K/L\rfloor\}$.} \label{tab_schemes}
           \resizebox{\textwidth}{!}{%
       \begin{tabular}{|c|c|c|c|}
    \hline
    Schemes & Load & Constraints & Subpacketization \\ \hline

    HKD scheme \cite{HKD} & $\frac{K-t L}{1+t}$& $L|K$  & $L\binom{K/L}{t}$   \\ \hline

    \tabincell{c}{RK1 scheme \cite{RK}}    & $\frac{(K-t L)^{2}}{K}$    &    & $\binom{K-t(L-1)-1}{t-1}\frac{K}{t}$ \\   \hline

    \tabincell{c}{CWLZC scheme \cite{CWLZC}}    & $\frac{K-t L}{1+t}$    &    & $\binom{K-t(L-1)}{t}K$ \\    \hline

    %RK2 scheme \cite{RK2}    & $\frac{\sum_{b \in B} \min \{2(K - t L) + t -1- \hat{b} , K\}}{|\hat{S}|(t + 1)}$ &    & $\binom{K-t(L-1)-1}{t-1}\frac{K}{t}$ \\\hline

    RK2 scheme~\cite{RK2} & $ \frac{K-tL}{K(t+1)|\mathcal B|} \sum_{\mathbf b\in\mathcal B} \min\{2(K-tL)+t-1-\|\mathbf b\|_\infty,K\}$ 
    & & $\binom{K-t(L-1)-1}{t-1}\frac{K}{t}$
\\\hline

 \multirow{3}{*}{WCWL scheme \cite{WCWL}}    & $\frac{(K-tL)(K-tL+1)}{2K}$    & $(K-tL+1)|K\ \text{or}\ K-tL=1$
     & $K$ \\\cline{2-4}& $\frac{K-tL}{2\lfloor\frac{K}{K-tL+1}\rfloor+1}$ & $\langle K\rangle_{K-tL+1}=K-tL$ and $K-tL>1$& $\left(2\lfloor\frac{K}{K-tL+1}\rfloor+1\right)K$ \\ \cline{2-4}& $\frac{K-tL}{2\lfloor\frac{K}{K-tL+1}\rfloor}$ & $\mbox{otherwise}$ & $2\lfloor\frac{K}{K-tL+1}\rfloor K$ \\ \hline

 \tabincell{c}{SPE scheme \cite{SPE}} & \tabincell{c}{$\geq \frac{K-2L}{4}$}  & $4|K(K-2L+2)$, $t=2$ & $\frac{K(K-2L+2)}{4}$ \\ \hline

 \multirow{3}{*}{SR1 scheme \cite{SR_arX2021}}& $\frac{1}{K}$  & $K - tL = 1$;  $t,K$ are coprime
     & \\ \cline{2-3} & $2\sum_{r=\frac{K-tL}{2}+1}^{K-tL}\frac{1}{1+\lceil\frac{tL}{r}\rceil}$
     & $K - tL$ is even; $t,K$ are coprime & \tabincell{c}{$\leq K^2$} \\  \cline{2-3} & $\frac{1}{\left(\lceil\frac{2tL}{K-tL+1}\rceil+1\right)} + \sum_{r=\frac{K-tL+3}{2}}^{K-tL}\frac{2}{1+\lceil\frac{tL}{r}\rceil}$  & $K - tL > 1$ is odd; \text{$t,K$ are coprime}
     & \\  \cline{2-3}  \hline

 \tabincell{c}{SR2 scheme \cite{SR}} & $\frac{(K-t L)(K-t L+t)}{2K}$ & $t|K$, $(K-t L+t)|K$
     & $K$ \\ \hline

 MR scheme \cite{MR}
     & $\geq\left\lceil \frac{K(K-L)}{2+2\lfloor \frac{L}{K-L+1} \rfloor} \right\rceil \frac{1}{K}$ & $t=1$ & $K$ \\ \hline

 LCWC scheme~\cite{LCWC} & $L\left\lfloor\frac{a-1}{2}\right\rfloor^n$ & $L\mid K,\ K=La^n,\ a\geq 3$ & $K$ \\ \hline
    
    \textbf{Scheme in Theorem~\ref{thm:main theorem}}& \boldmath{$\frac{(K-tL)(K-tL+t)}{K(t+1)}$}& & \boldmath{$\binom{K-t(L-1)-1}{t-1}\frac{K}{t}$} \\\hline
\end{tabular}}
\end{table*}

\subsection{Motivation and Contributions}
In this paper, we focus on the MACC problem under the cyclically $L$-separated placement, which avoids redundancy among the $L$ adjacent cache nodes accessed by any user, thereby achieving maximum local cache gain. This placement strategy is widely used in existing MACC coded caching schemes~\cite{HKD, RK, RK2, CWLZC}. For this placement strategy, the RK2 scheme~\cite{RK2} reduced the achievable transmission load by studying the structured index coding problem;  however, their scheme has not yet achieved the optimal transmission load under general parameters. The authors in~\cite{CWLZC}, on the other hand, established a lower bound on the transmission load for this fixed placement and proved that its achievable scheme is asymptotically optimal when $K\gg tL$. 
Neither of the two approaches above has succeeded in determining the upper and lower bounds of the transmission load for general system parameters; consequently, the optimal transmission load under this placement strategy has not yet been precisely characterized. 

Furthermore, exact optimality under fixed placement does not guarantee global optimality under general uncoded placement. While Reddy and Karamchandani~\cite{RK} established a lower bound on the transmission load for the full memory interval under the condition $L\geq K/2$, a unified performance characterization for arbitrary access $L$ is still lacking. 

Motivated by the above, this paper first establishes an exact converse bound on the transmission load for the cyclically $L$-separated placement strategy and presents an explicit linear coding caching scheme to achieve exact optimality under this placement. Furthermore, we propose a converse bound on the transmission load under uncoded placement, and prove the order optimality of the proposed scheme across the entire storage interval.

The main contributions of this paper are summarized as follows.
\begin{itemize}
    \item 
   Under the cyclically $L$-separated placement, we first propose a converse bound by using the acyclic index coding converse bound~\cite{ZYST, WTP2020}. Under this specific cache placement, a converse bound was proposed in~\cite[Theorem 2]{CWLZC}. 
   Even though both converses use the same placement and consecutive-user acyclic-bound framework, our proposed converse is strictly tighter; it follows from counting all valid cache supports in the remaining linear interval, including those omitted by the additional wrap-around separation constraint implicit in the numerical expression of \cite[Theorem~2]{CWLZC}.
    \item We then propose a MACC coded caching scheme with the cyclically $L$-separated placement and a new interference alignment strategy.  By constructing a random generating matrix $\mathbf{H}$ for the users (where each column of the matrix corresponds to a user), the proposed scheme broadcasts messages, each of which is a linear combination of the subfiles desired by all users; in each linear combination, the encoding coefficient of each subfile is the determinant of a sub-matrix of $\mathbf{H}$. Surprisingly,  by a unified approach to select the sub-matrix for each subfile, the interference of each user is aligned simultaneously into a limited dimension, while preserving the desired-signal dimension to be full and also separable from interference.\footnote{\label{foot:existing determinant coding}In the literature of coded caching, determinant-based encoding has been used in~\cite{Girgis2017} and determinant-based decoding has been used in~\cite{Wanscalar, Mascalar}. However, the proposed interference alignment approach, which uses determinant-based encoding, has significant differences compared to~\cite{Girgis2017}. First, the work in~\cite{Girgis2017} considered the cache-aided wireless interference networks with an arbitrary number of transmitters and receivers, while this paper considers  MACC systems. Second, the coded caching scheme in~\cite{Girgis2017} uses a cascade precoding approach, where the overall precoding matrix is the product of a determinant-based encoding matrix (by using the physical-channel cofactors) for zero forcing and using monomials of effective-channel coefficients (i.e., the product of physical channel and the precoding matrix for zero forcing) to align residual interference. However, in the proposed scheme, we design broadcast messages by using determinant-based encoding for interference alignment, while the zero forcing step is done in the decoding phase.}
    The proposed scheme is then proved to exactly match the above converse bound under the cyclically $L$-separated placement. 
    \item  Next, we propose a new converse bound under uncoded placement (not specifically for the cyclically $L$-separated placement), by averaging the acyclic index-coding converse bounds over cyclic user permutations. Comparing this converse bound and the proposed caching scheme, we show that the proposed scheme is generally order optimal within a factor of $2$ under uncoded cache placement.
    \item Finally, we provide theoretical and numerical comparisons with existing schemes. The proposed scheme achieves a lower transmission load than the RK1 and RK2 schemes~\cite{RK, RK2} with the same subpacketization. Compared with the CWLZC scheme~\cite{CWLZC}, our scheme reduces both the transmission load and subpacketization. It also achieves significant load reductions over several low-subpacketization schemes in~\cite{HKD, WCWL, LCWC, SR, SR_arX2021, SPE} with a moderate increase in subpacketization for selected system parameters.
\end{itemize}

\subsection{Organization and Notation}
The remainder of this paper is structured as follows. Section~\ref{sec:system_model} introduces the cyclical MACC system and the cyclically $L$-separated placement. Section~\ref{sec:Main_result} presents our main theoretical results and a performance analysis of existing schemes, including both theoretical and numerical comparisons. In Section~\ref{sec:fixed-placement-converse} and Section~\ref{sec:det-scheme}, we provide the detailed proof of the lower bound under the cyclically $L$-separated placement and the achievable coded caching scheme with an illustrative example, respectively. In Section~\ref{sec:global-converse}, we prove the global converse bound under uncoded placement. Finally, Section~\ref{sec:conclusion} concludes the paper. 

{Notations:} For a positive integer $a$, $[a]\triangleq\{1,2,\ldots,a\}$, and $[a:b]\triangleq\{a,a+1,\ldots,b\}$ for $a\leq b$. User and cache indices are interpreted modulo $K$ through $\langle i\rangle_K\in[K]$, with $\langle i\rangle_K=K$ when $K$ divides $i$. For a finite set $\mathcal{V}$, $|\mathcal{V}|$ is its cardinality; $(x)^+=\max\{x,0\}$. For a finite set $\mathcal{A}$, $\binom{\mathcal{A}}{b}$ is the family of its $b$-element subsets. For a matrix $\mathbf{H}$, $\mathbf{H}[\mathcal{A},\mathcal{B}]$ is the sub-matrix with row set $\mathcal{A}$ and column set $\mathcal{B}$, both ordered increasingly; a colon $:$ denotes all rows or all columns. Let $\mathbf{e}_i$ denote the column vector with a $1$ in the $i$-th entry and zeros elsewhere. The determinant of a square matrix $\mathbf{G}$ is denoted by $\det (\mathbf{G})$, and the determinant of the empty matrix is one. Let $\mathbb{F}_p$ denote the finite field with $p$ elements, where $p$ is a prime power. 
\section{System Model}\label{sec:system_model}

\subsection{The cyclical MACC system}\label{sbusec:MACC}
\begin{figure}
\centering
\centerline{\includegraphics[width=0.9\linewidth]{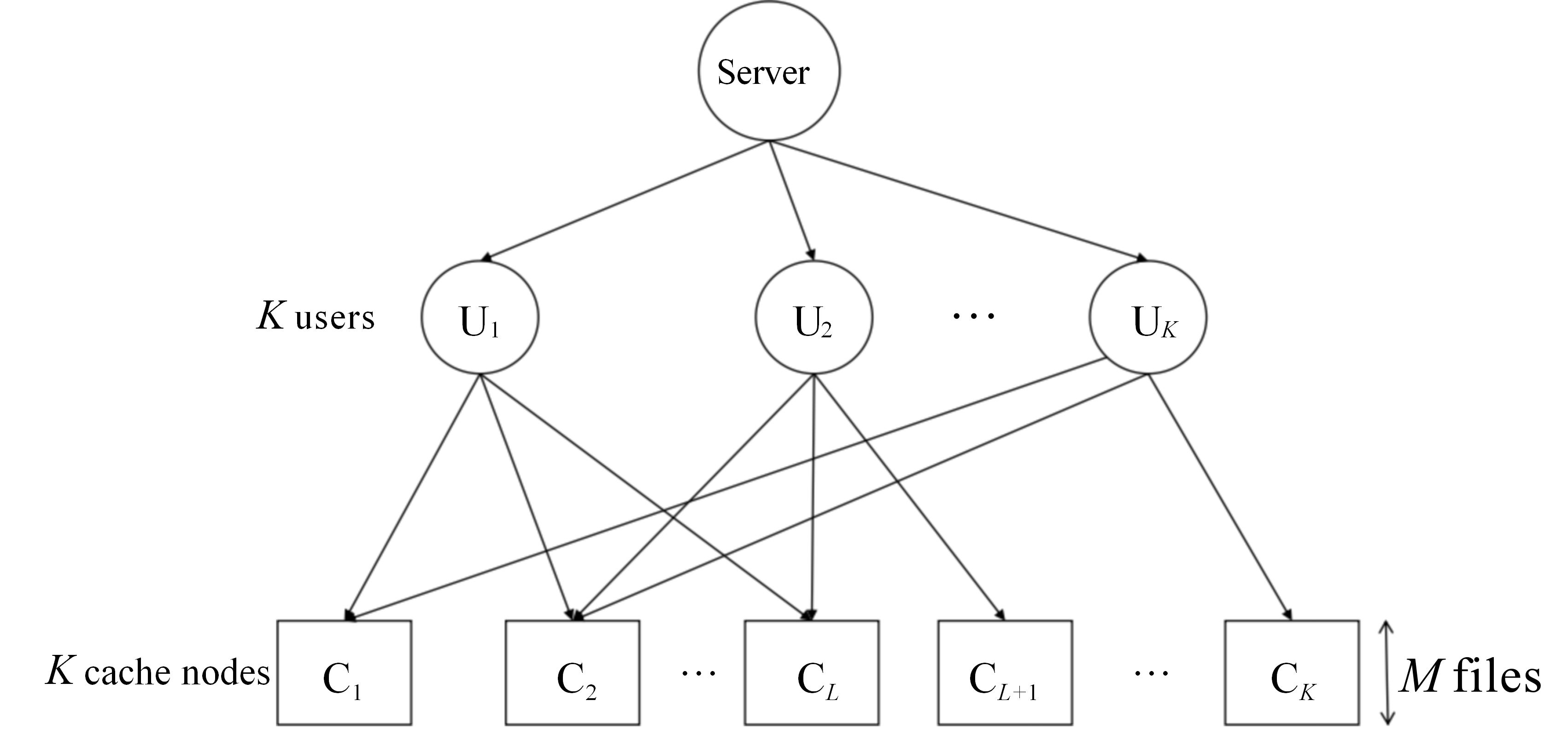}}
\caption{\small The $(K,L,M,N)$  multi-access  coded caching system.}
\label{fig:topology}
\end{figure}

We consider the $(K, L, M, N)$ cyclical MACC system proposed in~\cite{HKD}, which contains a central server, a library $\mathcal W=\{\mathbf{w}_1,\ldots, \mathbf{w}_N\}$ of mutually independent files (each with $B$ bits), \footnote{\label{foot:packet}%We assume that each packet has $B$ bits, such that our further coding scheme can be designed on the packet-level.  
The main focus of this paper is to characterize the optimal communication load, instead of the subpacketization; thus we assume that $B$ is large enough.} $K\leq N$ cache nodes C$_1$, $\ldots$, C$_K$, and $K$ cache-less users U$_1$, $\ldots$, U$_K$, as illustrated in Figure.~\ref{fig:topology}. Every cache has the capacity to cache at most $M$ files. The server communicates with all users through an error-free shared link, while user U$_k$ can also read the contents of the $L$ consecutive caches
\begin{equation}
 \mathcal A_k=\{\langle k\rangle_K,\langle k+1\rangle_K, \ldots,\langle k+L-1\rangle_K\}.
 \label{eq:access}
\end{equation}
As in the standard MACC model, cache access is assumed to incur no load on the shared link. 
A caching scheme contains the following two phases.

$\bullet$ {\bf Placement phase:} Each cache node C$_k$ where $k\in[K]$ stores at most $MB$ bits in its cache.
Let $\mathcal{Z}_k$ denote the content stored in cache node C$_k$. This paper considers uncoded cache placement, where each cache directly stores subsets of file bits. In that case, each file can be partitioned into subfiles as $\mathbf{w}_n=(\mathbf{w}_{n,\mathcal{T}}: n\in [N];\mathcal{T}\subseteq[K])$, where the subfile $\mathbf{w}_{n,\mathcal{T}}$ is stored exactly in the caches indexed by $\mathcal{T}$.
\iffalse
{\blue  For each user U$_k$ where $k\in[K]$, the placement constraints are
\begin{equation}
 x_{n,\mathcal{T}}\geq0,\qquad \sum_{\mathcal{T}\subseteq[K]}x_{n,\mathcal{T}}=1,\qquad
 \sum_{\mathcal{T}\subseteq [K], k\in \mathcal{T}}x_{\mathcal{T}}\leq\frac{M}{N},
 \label{eq:placement-constraints}
\end{equation}
which imply $\sum_\mathcal{T}|\mathcal{T}|x_\mathcal{T}\leq KM/N$. For a cache support $\mathcal{T}$, define the users that locally know $\mathbf{w}_{n,\mathcal{T}}$ and those that do not as
\begin{equation}
 \mathcal U(\mathcal{T})=\{k\in[K]:\mathcal A_k\cap \mathcal{T}\ne\emptyset\},  \qquad \bar{\mathcal U}(\mathcal{T})=[K]\setminus\mathcal U(\mathcal{T}).
 \label{eq:known-unknown}
\end{equation}
These sets determine the side-information graph induced by the placement.
} \fi

Let $x_{n,\mathcal{T}}=H(\mathbf{w}_{n,\mathcal{T}})/B$. For each cache support $\mathcal{T} \subseteq[K]$, define $x_{\mathcal{T}} \triangleq \frac{1}{NB}\sum_{n=1}^{N} H(\mathbf{w}_{n,\mathcal{T}})$. 
For each user U$_k$ where $k \in [K]$, the placement constraints are
\begin{equation}
x_{\mathcal{T}}\geq 0,\qquad \sum_{\mathcal{T}\subseteq[K]}x_{\mathcal{T}}=1,\qquad \sum_{\substack{\mathcal{T}\subseteq[K]\\k\in\mathcal{T}}}
x_{\mathcal{T}} \leq \frac{M}{N}, \quad k\in[K].
\label{eq:placement-constraints}
\end{equation}
Summing the last inequality over all $k\in[K]$ gives $\sum_{\mathcal{T}}
|\mathcal{T}|x_{\mathcal{T}} \leq KM/N$. For a cache support $\mathcal{T}$, define the users that locally know $\mathbf{w}_{n,\mathcal{T}}$ and those that do not as
\begin{equation}
 \mathcal U(\mathcal{T})=\{k\in[K]:\mathcal A_k\cap \mathcal{T}\ne\emptyset\},  \qquad \bar{\mathcal U}(\mathcal{T})=[K]\setminus\mathcal U(\mathcal{T}).
 \label{eq:known-unknown}
\end{equation}
These sets determine the side-information graph induced by the placement.

$\bullet$ {\bf Delivery phase:} Each user requests one file from the library. Let $\mathbf{d}=(d_1,\ldots,d_K)\in[N]^K$ denote the demand vector, where user U$_k$ requests file $\mathbf{w}_{d_k}$. Given $\mathbf{d}$, the server broadcasts $\mathbf{X}_{\mathbf{d}}=\psi_{\mathbf{d}}(\mathcal{W})$ over the shared link. User U$_k$ must satisfy
\begin{equation}
 H(\mathbf{w}_{d_k}\mid \mathbf{X}_{\mathbf{d}},(\mathcal{Z}_{k'}:{k'}\in\mathcal A_k),
       \mathbf{d})=0,\qquad k\in[K].
 \label{eq:recovery}
\end{equation}
A memory-load pair $(M, R)$ is achievable if~\eqref{eq:recovery} holds for every demand vector and $H(\mathbf{X}_{\mathbf{d}})\leq RB$. The optimal worst-case load under uncoded placement is defined as $ R^{\star}_{\rm u}$.

\begin{remark}
\label{rem:zero-load point}
    Note that for the zero-load point (i.e., $R=0$), it has been shown in~\cite{HKD} that the minimum required memory is $N/L$, by using the Maximum Distance Separable (MDS) code in the placement phase such that each user can obtain all the files from its connected $L$ cache nodes. Accordingly, the coded cache placement is only considered for this zero-load point $(M, R)=(N/L,0)$ in the rest of the paper (i.e., we consider the lower convex envelope of the memory-load tradeoff points by schemes with uncoded cache placement and the point $(M, R)=(N/L,0)$ achieved by the MDS code in the placement phase). 
    \hfill $\square$ 
\end{remark}

\subsection{The cyclically \texorpdfstring{$L$}{L}-separated placement}
\label{sub:Lseparate placement}
For the cyclical MACC system, in order to achieve the maximum local caching gain at the users (i.e., each user can retrieve $LMB$ bits from its connected caches),  any $L$ neighbouring cache nodes do not cache any common packet. A cyclically $L$-separated placement has been widely used in the existing works on the cyclical MACC system~\cite{HKD, RK, RK2, CWLZC}. This placement can be described as follows.

 Consider each memory point $M/N=t/K$ where $ t\in \left[0: \left\lfloor K/L \right\rfloor \right]$. \footnote{\label{foot:trivial case}  When $t=0$, we have $M=0$, all caches are empty. Under a worst-case demand vector consisting of $K$ distinct requests, which exists since $N\ge K$, the server must transmit $K$ file units; this load is achieved by uncoded transmission of the $K$ requested files. Hence, the optimal load and the corresponding converse bound are both equal to $K$. Hence, we consider $t\ge 1$ in the remainder of the paper and treat $t=0$ separately.}  
 Define $q\triangleq K-tL$. Each file $\mathbf{w}_n$ where $n\in [N]$ is divided into $\binom{q+t-1}{t-1}\frac{K}{t}$ non-overlapping and equal-length subfiles, where
\begin{align}
\label{eq:our subfile division}
\mathbf{w}_n=\{\mathbf{w}_{n,\mathcal{T}}: \mathcal{T} \subseteq [K], |\mathcal{T}|=t, L\leq \langle j_1-j_2\rangle_K, \forall j_1,j_2 \in \mathcal{T} \text{ and } j_1 \neq j_2  \}.
\end{align}
Hence, each subfile is cached by $t$ cache nodes, and any two cache nodes within distance $L-1$ do not cache any common packets. For notational simplicity, we define 
 \begin{align}
 \mathfrak{T}:= \{\mathcal{T} \subseteq [K]: |\mathcal{T}|=t, L\leq \langle j_1-j_2\rangle_K, \forall j_1,j_2 \in \mathcal{T} \text{ and } j_1 \neq j_2  \}. \label{eq:St}
 \end{align}
  Thus $\mathbf{w}_n=\{\mathbf{w}_{n,\mathcal{T}}: \mathcal{T} \in \mathfrak{T}\}$. Note that for other memory points, memory sharing between the above points and $(M, R)=(N/L,0)$ can be used. The optimal worst-case load under the cyclically $L$-separated placement is defined as  $R^{\star}_{\rm cyc}$.

\section{Main Results and Performance Analysis}\label{sec:Main_result}
This section presents the main results and analyzes the transmission load of the proposed scheme. We first present a converse bound under cyclically $L$-separated placement and a linear delivery scheme that attains this bound. We then derive a converse bound under arbitrary uncoded placement and establish the order optimality of the proposed scheme within this class. Finally, we compare its transmission load with those of existing schemes.

Fixing the cyclically $L$-separated placement, we characterize the optimal worst-case load in the following theorem, where we propose a new converse bound by extending the acyclic index coding bound~\cite{WTP2020} and a new achievable scheme by a determinant-based interference alignment design.  The proof of the converse bound is provided in Section~\ref{sec:fixed-placement-converse} and the description of the proposed scheme is provided in Section~\ref{sec:det-scheme}.
\begin{theorem}[Optimal load under the cyclically $L$-separated placement]
\label{thm:main theorem}
 For the $(K, L, M, N )$ cyclical MACC system with $N\geq K$, under the  cyclically $L$-separated placement, the optimal worst-case load-memory tradeoff points are 
\begin{align}
    (M, R^{\star}_{\rm cyc})=\left( \frac{N t}{K} ,\frac{q(q+t)}{K(t+1)} \right), \ \forall t\in \left[0: \left\lfloor\frac{K}{L} \right\rfloor \right].
    \label{eq:optimal load under fixed placement}
\end{align}
\hfill $\square$
\end{theorem}

Define 
\begin{align}\label{eq_Phi_t}
    \Phi_t :=\frac1K\sum_{i=0}^{K-1}\bigl(K- t(L+i)\bigr)^+,
\end{align}
for each $t\in \left[0: \left\lceil \frac{K}{L} \right\rceil \right]$. Next we propose a converse bound for the general uncoded cache placement, whose proof is provided in Section~\ref{sec:global-converse}.
\begin{theorem}[Converse bound under uncoded cache placement]\rm
\label{thm:uncoded converse}
 For the $(K, L, M, N )$ cyclical MACC system with $N\geq K$, under uncoded cache placement,
\begin{equation}
 R^{\star}_{\rm u} \geq  {\rm Conv}\left(\left\{\left(\frac{Nt}{K}, \Phi_t\right): t\in \left[0: \left\lceil K/L \right\rceil \right]\right\}\right),
 \label{eq:global-bounds}
\end{equation}
where ${\rm Conv}\left(\left\{(\frac{Nt}{K}, \Phi_t): t\in \left[0: \left\lceil K/L\right\rceil \right]\right\}\right)$ represents the convex lower envelope of the points  $(\frac{Nt}{K}, \Phi_t)$ for all $t\in \left[0: \left\lceil \frac{K}{L} \right\rceil \right]$.
\hfill $\square$
\end{theorem}

Comparing the converse bound under uncoded cache placement in Theorem~\ref{thm:uncoded converse} and the achievable scheme of  Theorem~\ref{thm:main theorem}, we can obtain the following order optimality result, whose proof is provided in Appendix~\ref{app:uncoded-two-approx}.
\begin{theorem}[Order optimality under uncoded cache placement]\rm
\label{thm:uncoded-two-approx}
 For the $(K, L, M, N )$ cyclical MACC system with $N\geq K$,
the achievable load-memory tradeoff of  Theorem~\ref{thm:main theorem} is order optimal within a factor of $2$ under uncoded cache placement.
\hfill $\square$
\end{theorem}
 A more refined comparison of the achievability result of Theorem~\ref{thm:main theorem} and the converse bound of Theorem~\ref{thm:uncoded converse} reveals the tight optimality of the scheme of Theorem~\ref{thm:main theorem}  for the first segment of the load-memory tradeoff (i.e., when the cache memory is small). In particular, this reveals that for this regime the $L$-separated placement is optimal over all uncoded placement schemes, as formally proved in Appendix~\ref{sec:proof of cor}. 
 \begin{corollary}[Exact first memory segment]\rm
 \label{cor:optimal regime}
For the $(K, L, M, N )$ cyclical MACC system with $N\geq K$ and $0\leq M\leq N/K$,
\begin{equation}
 R_{\mathrm{unc}}^{\star}(M)=K-\frac{KM}{N}
 \left(K-\frac{(K-L)(K-L+1)}{2K}\right).
 \label{eq:first-segment}
\end{equation}
\hfill $\square$
\end{corollary}

Finally, we compare the transmission load of the scheme of Theorem~\ref{thm:main theorem} with those of the schemes summarized in Table~\ref{tab_schemes}. Under the corresponding parameter constraints, the transmission loads of HKD, RK1, CWLZC, SR1, and the listed LCWC construction~\cite{HKD, RK, CWLZC, SR_arX2021, LCWC} are no smaller than that of RK2, while those of SR2 and MR~\cite{SR, MR} are no smaller than that of WCWL. Therefore, it suffices to compare the proposed scheme with the RK2~\cite{RK2}, SPE~\cite{SPE}, and  WCWL~\cite{WCWL} schemes.
\subsection{Theoretical Comparisons}
\label{sub:theoretic performance}

$\bullet$ {\bf Comparison with the RK2 scheme~\cite{RK2}:} Let $N\geq K$, $L>1$, $M/N=t/K$,
$t\in\{1,\ldots,\lfloor K/L\rfloor\}$, and $q=K-tL>0$.
The scheme of Theorem~\ref{thm:main theorem} achieves
\begin{equation}
 R_{\rm cyc}^{\star}=\frac{q(q+t)}{K(t+1)}.
 \label{eq:comparison-proposed-load}
\end{equation}
Define
\[
 \mathcal B=\left\{\mathbf b\in\mathbb Z_{\geq0}^{t+1}:
            \sum_{i=1}^{t+1}b_i=q-1\right\}, \qquad
 |\mathcal B|=\binom{q+t-1}{t}.
\]
The load of RK2~\cite{RK2} can be expressed as
\begin{equation}
 R_{\rm RK2}
 =\frac{q}{K(t+1)|\mathcal B|}
   \sum_{\mathbf b\in\mathcal B}
       \min\{2q+t-1-\|\mathbf b\|_\infty,K\}.
 \label{eq:comparison-rk2-load}
\end{equation}
Since $\|\mathbf b\|_\infty\leq q-1$ and $K=q+tL\geq q+t$,
each summand is at least $q+t$. Therefore,
\begin{equation}
 \frac{R_{\rm cyc}^{\star}}{R_{\rm RK2}}
 =\frac{(q+t)|\mathcal B|}
       {\displaystyle\sum_{\mathbf b\in\mathcal B}
        \min\{2q+t-1-\|\mathbf b\|_\infty,K\}}
 \leq1.
 \label{eq:comparison-rk2-ratio}
\end{equation}
If $q\geq3$, the vector
$\mathbf b=(q-2,1,0,\ldots,0)$ gives a summand equal to
$q+t+1$, so the inequality is strict. Equality holds when
$q\in\{1,2\}$.
The proposed scheme and RK2 use the same placement and have the same
subpacketization, so this load reduction does not require further
file splitting.

$\bullet$ {\bf Comparison with the SPE scheme~\cite{SPE}:} 
For the $t=2$ construction of SPE~\cite{SPE}, each transmitted XOR serves at most four missing packets. Its load therefore satisfies
$R_{\rm SPE}\geq q/4$, and
\begin{equation}
 \frac{R_{\rm cyc}^{\star}}{R_{\rm SPE}}
 \leq\frac{4(q+2)}{3K}.
 \label{eq:comparison-spe-ratio}
\end{equation}
Thus, the proposed load is no greater than the SPE load when $K\leq8L-8$, with strict improvement when $K<8L-8$, provided that the SPE construction is applicable. Outside this range, the estimate alone does not determine which scheme has the smaller load. The numerical comparison below uses the actual SPE load from~\cite[eqs.~(2) and~(3)]{SPE}.

$\bullet$ {\bf Comparison with the WCWL scheme~\cite{WCWL}: }
The three load expressions of WCWL~\cite{WCWL} in Table~\ref{tab_schemes} all satisfy
\begin{equation}
 R_{\rm WCWL}\geq\frac{q(q+1)}{2K}.
 \label{eq:comparison-wcwl-lower-estimate}
\end{equation}
Indeed, the denominator in each expression, when the numerator is written as $q$, is at most $2K/(q+1)$.
Consequently,
\begin{equation}
 \frac{R_{\rm cyc}^{\star}}{R_{\rm WCWL}} \leq\frac{2(q+t)}{(t+1)(q+1)} =1-\frac{(t-1)(q-1)}{(t+1)(q+1)} \leq1.
 \label{eq:comparison-wcwl-ratio}
\end{equation}
The reduction is strict for $t>1$ and $q>1$. Equality holds precisely when $q=1$, or when $t=1$ and $(q+1)\mid K$.
\subsection{Numerical Comparisons}
\label{sub:numerical performance}
We consider $K=20$, $L=4$, and $N \geq K$. For the RK2 and WCWL schemes, we evaluate their transmission load at the integer memory points $M/N=t/20$ for $1\leq t\leq 4$. For the SPE scheme, the applicable point $M/N=2/20$ has load $R_{\rm SPE}=23/7$, obtained by evaluating the exact expression in~\cite{SPE}. Each scheme is supplemented by $(M/N,R)=(0,K)$ and $(1/L,0)$. Intermediate points are achieved by memory sharing. Since $L\mid K$ in this example, the zero-load point is achievable under uncoded placement: divide each file into $L$ equal parts and repeat these parts periodically across the caches, so that any $L$ consecutive caches contain the entire library.

\begin{figure}[!tbp]
 \centering
 \includegraphics[width=0.8\linewidth]{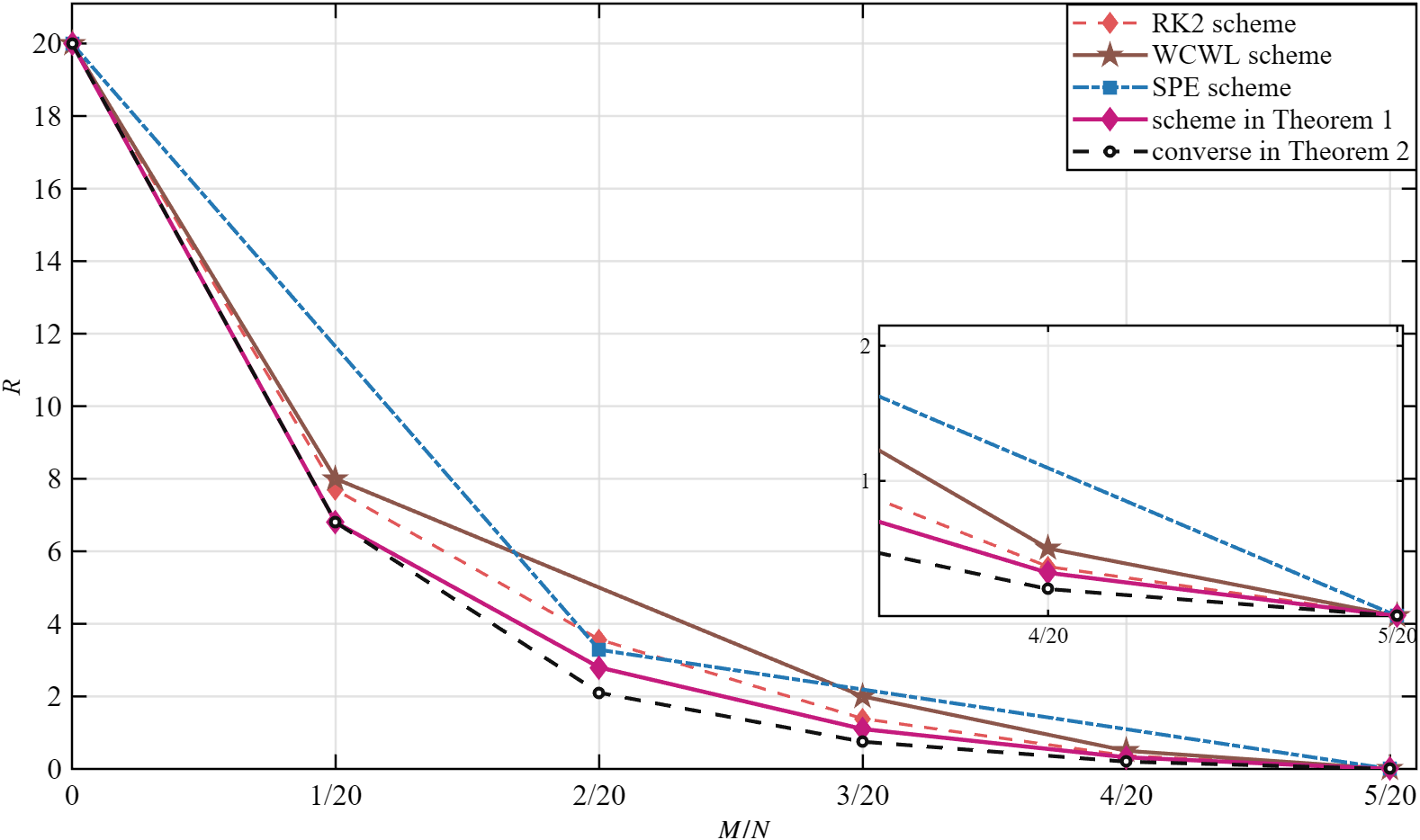}
 \caption{Transmission load versus cache-node memory ratio for
 $K=20$ and $L=4$. The best reference curve is the lower convex
 envelope of the combined achievable points of RK2, WCWL, and
 SPE. All achievable curves use uncoded placement.}
 \label{fig:comparison-load}
\end{figure}

Figure~\ref{fig:comparison-load} shows the loads of the three reference schemes, the proposed load, and the converse in Theorem~\ref{thm:uncoded converse}. The best reference curve is obtained by taking the lower convex
envelope of their combined achievable points, allowing memory sharing across the three reference schemes.
For this example, the vertices of the combined reference curve are attained by RK2 at $t=1,3,4$ and by SPE at $t=2$, in addition to the two common endpoints. The proposed load is strictly lower than this best reference
curve for $0<M/N<1/4$ and coincides with it at the endpoints. For example, at $M/N=3/20$, the smallest reference load is $104/75$, whereas the proposed scheme achieves $11/10$, a reduction of approximately $20.7\%$.

The proposed load coincides with the uncoded-placement converse for $0\leq M/N\leq1/20$, establishing optimality over this memory
interval. For $1/20<M/N<1/4$, the ratio of the proposed load to the converse is at most $8/5$ in this example, certifying a multiplicative gap of at most $8/5$ to the optimal load under arbitrary uncoded placement.
At $M/N=1/4$, the achievable load and the converse are both zero.

\paragraph*{Comparison with all schemes in Table~\ref{tab_schemes}}
Table~\ref{tab:all-scheme-loads} gives the exact achievable loads at selected original parameter points of all the schemes in Table~\ref{tab_schemes}. Each column uses the same $(K,L,t)$ and memory ratio $M/N=t/K$ for all applicable schemes, without memory sharing. These points include at least one applicable point for every reference scheme. The point $(12,4,2)$ includes SR2, which has no positive-load integer point for $K=20$ and $L=4$. The LCWC entry uses $a=3$ and $n=2$, giving $(K,L,t)=(27,3,5)$. The SPE and MR entries use their actual achievable loads.

At every applicable point shown, the RK2 load is no greater than
those of HKD, RK1, CWLZC, SR1, and LCWC, while the WCWL load
is no greater than those of SR2 and MR. These values agree with
the load relations used above to select the three reference
schemes. The proposed scheme achieves a strictly smaller load
than every applicable reference scheme in this table.

\begin{table}[!tbp]
 \centering
 \small
 \renewcommand{\arraystretch}{1.25}
 \setlength{\tabcolsep}{5pt}
 \caption{Achievable transmission loads of all schemes in
 Table~\ref{tab_schemes} at selected original parameter points.
 Each column specifies $(K,L,t)$, with $M/N=t/K$ and $N\geq K$.
 A dash denotes an inapplicable original parameter point.
 No memory sharing is applied.}
 \label{tab:all-scheme-loads}
 \begin{tabular}{lccccc}
 \hline
 Scheme & $(20,4,1)$ & $(20,4,2)$ & $(20,4,3)$
        & $(12,4,2)$ & $(27,3,5)$ \\
 \hline
 HKD~\cite{HKD} & $8$ & $4$ & $2$ & $4/3$ & $2$ \\
 RK1~\cite{RK} & $64/5$ & $36/5$ & $16/5$ & $4/3$ & $16/3$ \\
 CWLZC~\cite{CWLZC} & $8$ & $4$ & $2$ & $4/3$ & $2$ \\
 RK2~\cite{RK2} & $77/10$ & $232/65$ & $104/75$ & $34/45$ & $320/189$ \\
 WCWL~\cite{WCWL} & $8$ & $6$ & $2$ & $1$ & $3$ \\
 SPE~\cite{SPE} & --- & $23/7$ & --- & $1$ & --- \\
 SR1~\cite{SR_arX2021} & $8$ & --- & $5/2$ & --- & $23/6$ \\
 SR2~\cite{SR} & --- & --- & --- & $1$ & --- \\
 MR~\cite{MR} & $8$ & --- & --- & --- & --- \\
 LCWC~\cite{LCWC} & --- & --- & --- & --- & $3$ \\
 \hline
 Proposed & $34/5$ & $14/5$ & $11/10$ & $2/3$ & $34/27$ \\
 \hline
 \end{tabular}
\end{table}

\section{Proof of Theorem~\texorpdfstring{\ref{thm:main theorem}}{1}: Converse Bound Under Cyclically \texorpdfstring{$L$}{L}-separated Placement}
\label{sec:fixed-placement-converse}

We prove the converse part of Theorem~\ref{thm:main theorem}
at each integer memory point $M=Nt/K$, where
$t\in\{0,\ldots,\lfloor K/L\rfloor\}$ and $N\geq K$.
The placement is the cyclically $L$-separated placement defined
in Section~\ref{sub:Lseparate placement}.
Set $q=K-tL$.

When $t=0$, all caches are empty. A distinct-demand vector
requires the transmission of $K$ independent files, giving
$R\geq K$. When $q=0$, the desired converse is $R\geq0$,
which is immediate. Hence, we consider $t\geq1$ and $q\geq1$
in the following.

Recall that $\mathfrak T$, defined in~\eqref{eq:St}, is the family of all cyclically $L$-separated $t$-subsets of $[K]$, with $|\mathfrak T|  =\frac{K}{t}\binom{q+t-1}{t-1}$. 
Each file $\mathbf w_n$ is partitioned into $|\mathfrak T|$ equal-length subfiles
$\mathbf w_{n,\mathcal T}$, $\mathcal T\in\mathfrak T$. Subfile $\mathbf w_{n,\mathcal T}$ is stored in cache node C$_{\lambda}$ if and only if $\lambda\in\mathcal T$.
Since each file contains $B$ bits,
\begin{equation}
 H(\mathbf w_{n,\mathcal T})  =\frac{B}{|\mathfrak T|}, \qquad n\in[N],\quad \mathcal T\in\mathfrak T.
 \label{eq:integer-converse-subfile-size}
\end{equation}

Fix a demand vector $\mathbf d=(d_1,d_2,\dots,d_K)\in[N]^K$ with distinct entries, which exists because $N\geq K$. Following~\cite{ZYST, WTP2020}, for the demand vector $\mathbf d$, we construct a directed graph $\mathcal{G}_{\mathbf d}$,  where the vertex set is defined as
\begin{equation}
\mathfrak V= \left\{ \mathbf w_{d_k,\mathcal T}:
k\in[K],\; \mathcal T\in\mathfrak T,\; \mathcal T\cap\mathcal A_k=\emptyset
\right\}.
\label{eq:integer-converse-full-vertex-set}
\end{equation}
That is, each vertex represents a requested subfile that cannot be retrieved from its requesting user's accessible caches. A directed edge from a subfile requested by user U$_k$ to one requested by user U$_{k'}$ indicates that user U$_k$ knows the latter subfile from its accessible caches.

Consider $K$ users U$_1$, U$_2,\dots,$ U$_K$  ordered as $(1,2,\ldots,K)$. For $i\in[q]$, the caches accessible to the first $i$ users form the nonwrapping interval
\begin{equation}
 \bigcup_{j=1}^{i}\mathcal A_j=[1:L+i-1].
 \label{eq:integer-converse-cache-prefix}
\end{equation}
Indeed, $i\leq q=K-tL$ implies $L+i-1\leq K-(t-1)L-1<K$. For each $i\in[q]$, retain the subfiles requested by user U$_i$ that are unknown to all of the first $i$ users:
\begin{equation}
 \begin{aligned}
 \mathcal V_i
 =\left\{
    \mathbf w_{d_i,\mathcal T}:
    \mathcal T\in\mathfrak T,\;
    \mathcal T\cap[1:L+i-1]=\emptyset
   \right\},\qquad 
 \mathcal V=\bigcup_{i=1}^{q}\mathcal V_i.
 \end{aligned}
 \label{eq:integer-converse-selected-subfiles}
\end{equation}
Since $\mathcal A_i\subseteq[1:L+i-1]$, every subfile in $\mathcal V_i$ is unknown to its requesting user U$_i$. Consequently, $\mathcal V_i\subseteq\mathcal W_{\mathbf d}$ for every $i\in[q]$, and hence $\mathcal V\subseteq\mathfrak  V$. We now consider the subgraph of $\mathcal G_{\mathbf d}$ induced by $\mathcal V$.

Each subfile in $\mathcal V_j$ is unknown to users U$_1,$ U$_2,\dots$, U$_j$. Thus, every directed edge within $\mathcal V$ goes from a higher-indexed level to a lower-indexed level, and the induced subgraph is acyclic.
The acyclic index-coding bound~\cite{ZYST,WTP2020} yields
\begin{equation}
 \begin{aligned}
 H(\mathbf X_{\mathbf d})
\geq
 \sum_{i=1}^{q}
 \sum_{\mathbf w_{d_i,\mathcal T}\in\mathcal V_i}
 H(\mathbf w_{d_i,\mathcal T})=\frac{B}{|\mathfrak T|}
   \sum_{i=1}^{q}|\mathcal V_i|.
 \end{aligned}
 \label{eq:integer-converse-entropy}
\end{equation}
Since $RB\geq H(\mathbf X_{\mathbf d})$, we obtain
\begin{equation}
 R\geq\frac{1}{|\mathfrak T|}
       \sum_{i=1}^{q}|\mathcal V_i|.
 \label{eq:integer-converse-counting-bound}
\end{equation}

It remains to count $|\mathcal V_i|$. A subfile belongs to $\mathcal V_i$ if and only if its support
$\mathcal T=\{\lambda_1,\ldots,\lambda_t\}$ satisfies
\[
 L+i\leq \lambda_1<\cdots<\lambda_t\leq K,
 \qquad
 \lambda_{j+1}-\lambda_j\geq L,\quad j\in[t-1].
\]
The cyclic closing-gap constraint is automatic because $K+\lambda_1-\lambda_t\geq L+i>L$. Thus, the counting problem is on a path. Replacing $K$ by $K-(L+i-1)$ in the cyclic counting formula would incorrectly impose an additional closing-gap constraint.

For each $j\in[t]$, define the scalar  $\ell_j=\lambda_j-(L+i-1)-(j-1)(L-1)$. This transformation bijectively maps the admissible supports to the increasing tuples
\[
 1\leq \ell_1<\cdots<\ell_t \leq K-(L+i-1)-(t-1)(L-1)=q+t-i.
\]
Consequently,
\begin{equation}
 |\mathcal V_i|=\binom{q+t-i}{t}, \qquad i\in[q].
 \label{eq:integer-converse-prefix-count}
\end{equation}
Summing these counts and applying the hockey-stick identity gives
\begin{equation}
 \begin{aligned}
 \sum_{i=1}^{q}|\mathcal V_i|
 =\sum_{i=1}^{q}\binom{q+t-i}{t}=\sum_{j=t}^{q+t-1}\binom{j}{t}
 =\binom{q+t}{t+1}.
 \end{aligned}
 \label{eq:integer-converse-total-count}
\end{equation}
Substituting~\eqref{eq:integer-converse-total-count}
into~\eqref{eq:integer-converse-counting-bound}, we obtain
\begin{equation}
 R\geq
 \frac{\binom{q+t}{t+1}}
      {\frac{K}{t}\binom{q+t-1}{t-1}}
 =\frac{q(q+t)}{K(t+1)}.
 \label{eq:integer-converse-final-bound}
\end{equation}
This proves the converse in Theorem~\ref{thm:main theorem}.

\section{Proof of Theorem~\ref{thm:main theorem}: Proposed Scheme}
\label{sec:det-scheme}
 In this section, we propose a linear delivery scheme based on interference alignment, in which each broadcast packet is an explicit sum of file packets with determinant coefficients.  More precisely, based on the cyclically $L$-separated placement,   we construct the broadcast messages (each of which is a linear combination of all subfiles desired by all users with determinant-based coefficients),\footnote{\label{foot:elementwise linear combination}Since the length of each subfile is large enough, we can first map each subfile into a large enough finite field ${\mathbb F}_{p}$; then element-wise linear combinations can be done over ${\mathbb F}_{p}$.
}
cancel all unknown interference by linear combinations of these packets,
and use the Schwartz--Zippel lemma to show the decodability of the scheme. 

Throughout this section, recall that $q=K-tL$ and $1\leq t\leq\lfloor K/L\rfloor$. The family $\mathfrak{T}$ consists of all cyclically $L$-separated cache supports of size $t$. The index $\mathcal{T}$ in $\mathbf{w}_{n,\mathcal{T}}$ denotes the \emph{cache support}, rather than a set of users. For such a support, let $\mathcal U(\mathcal{T})$ and $\bar{\mathcal U}(\mathcal{T})$ denote the users that locally know and do not locally know $\mathbf{w}_{n,\mathcal{T}}$, respectively. All matrix subindices are listed in increasing order unless an ordered
column list is explicitly specified. We use the convention that the determinant of the empty matrix equals one.

This section is organized as follows.  We first describe the general delivery scheme in Section~\ref{sub:general encoding rule}.  Then Section~\ref{subsec:sv-example}  illustrates the main idea of the proposed scheme through an example where  $K=N=6$, $t=2$, and $L=2$. 

\subsection{General Determinant-Based Caching Scheme}
\label{sub:general encoding rule}
Consider $N\geq K$, $1\leq L\leq K$, and an integer $1\leq t\leq\lfloor K/L\rfloor$. Set $q=K-tL$ and $M=Nt/K$. We describe the main construction for $q\geq2$ and handle the boundary cases separately below.

\subsubsection{Placement and the packets missing at each user}
Recall that $\mathfrak{T}$ is the family of cyclically $L$-separated
$t$-subsets of $[K]$, where $F=|\mathfrak{T}|=\frac Kt\binom{q+t-1}{t-1}.$

We partition each file $\mathbf{w}_n$ where $n\in [N]$ into $|\mathfrak{T}|$ equal-length packets $\mathbf{w}_{n,\mathcal{T}}$,
$\mathcal{T}\in\mathfrak{T}$, and store $\mathbf{w}_{n,\mathcal{T}}$ in cache node C$_{\lambda}$  when
$\lambda\in \mathcal{T}$. 

Define the locally informed and uninformed users of a support by
\begin{equation}
 \mathcal U(\mathcal{T})=\{k:\mathcal A_k\cap \mathcal{T}\neq\emptyset\},
 \qquad
 \bar{\mathcal U}(\mathcal{T})=[K]\setminus\mathcal U(\mathcal{T}).
 \label{eq:sv-UQ}
\end{equation}
Cache node C$_{\lambda}$ informs the $L$ users U$_u$, where $u \in \{\langle \lambda-L+1\rangle_K,\langle \lambda-L+2\rangle_K,\ldots,\langle \lambda\rangle_K\}$. The $t$ such
intervals associated with $\mathcal{T}$ are disjoint: an overlap would place
two selected caches inside one user's length-$L$ access interval,
contrary to separation. So we have
\begin{equation}
 |\mathcal U(\mathcal{T})|=tL,\qquad |\bar{\mathcal U}(\mathcal{T})|=q.
 \label{eq:det-known-missing}
\end{equation}
In addition, for each user U$_k$ such that $k\in\bar{\mathcal U}(\mathcal{T})$, define
\begin{align}
 \mathcal D_{k,\mathcal{T}}=\bar{\mathcal U}(\mathcal{T})\setminus\{k\},
 \label{eq:det-other-missing}
\end{align}
where $|\mathcal D_{k,\mathcal{T}}|=q-1$.
For user U$_k$, let
\begin{equation}
 \mathcal E_k=\{\mathcal{T}\in\mathfrak{T}:k\in\bar{\mathcal U}(\mathcal{T})\};
 \label{eq:sv-E}
\end{equation}
in other words, user U$_k$ cannot retrieve $\mathbf{w}_{i,\mathcal{T}}$ from its connected caches, where $i\in [N]$  and $\mathcal{T} \in  \mathcal E_k$. 
Thus, each user needs
\begin{equation}
 |\mathcal E_k|=\frac{Fq}{K}
  =\frac qt\binom{q+t-1}{t-1}
  =\binom{q+t-1}{t}=\binom{q+t-1}{q-1}
 \label{eq:det-missing-count}
\end{equation}
subfiles of its requested file.

\subsubsection{Delivery phase}
The demand vector  $\mathbf{d}\in[N]^K$ is revealed. 
A key point of this scheme is to construct a generating matrix $\mathbf{H} \in {\mathbb F}_{p}^{(q+t)\times K}$, where each column corresponds to one user.  More precisely,  we let
\begin{align}
\mathbf{H}=[\mathbf{h}_1, \cdots, \mathbf{h}_K]\in\mathbb F_p^{(q+t)\times K};
 \label{eq:sv-iid-H}
\end{align}
 the element located at the $a$-th row and $k$-th column of $\mathbf{H}$ is denoted as $ h_{a,k}$, which is selected uniformly i.i.d. over ${\mathbb F}_{p}$.

For each $\mathcal{Q}\in\binom{[q+t]}{q-1}$, the server broadcasts a message which is a linear combination of all subfiles required by the users, 

\begin{equation}
\mathbf{x}_{\mathcal{Q}}=\sum_{k=1}^{K}\sum_{\substack{\mathcal{T}\in\mathfrak{T}:\\k\in\bar{\mathcal U}(\mathcal{T})}}
       \det \mathbf{H}[\mathcal{Q},\mathcal D_{k,\mathcal{T}}]\,\mathbf{w}_{d_k,\mathcal{T}}.
 \label{eq:det-broadcast}
\end{equation}
Note that in~\eqref{eq:det-broadcast}, the label $\mathcal{Q}$ selects coordinate rows of $\mathbf{H}$. Each coefficient is the
determinant of a $(q-1)\times(q-1)$ matrix. The number of broadcast
messages is
$
 \binom{q+t}{q-1}=\binom{q+t}{t+1},
$
where each message has $\frac{1}{|\mathfrak{T}|}= \frac{1}{\frac{K}{t}\binom{q+t-1}{t-1}}$ file units.  Hence, the achieved load is 
$\frac{\binom{q+t}{t+1}}{\frac Kt\binom{q+t-1}{t-1}}
 =\frac{q(q+t)}{K(t+1)}$ 
 which coincides with~\eqref{eq:optimal load under fixed placement}.

\subsubsection{Decoding process}
Let us consider user U$_k$ where $k\in [K]$. For each $\mathcal{Q}\in\binom{[q+t]}{q-1}$, user U$_k$ first removes the subfiles that are cached by its connected caches from $\mathbf{x}_{\mathcal{Q}}$ to obtain 
\begin{subequations}
\begin{align}
\mathbf{y}_{k,\mathcal{Q}}&=\mathbf{x}_{\mathcal{Q}}
 -\sum_{k'=1}^{K}
  \sum_{\substack{\mathcal{T}\in\mathfrak{T}:\\
                   k'\in\bar{\mathcal U}(\mathcal{T}),\ k\notin\bar{\mathcal U}(\mathcal{T})}}
     \det \mathbf{H}[\mathcal{Q},\mathcal D_{k',\mathcal{T}}]\,\mathbf{w}_{d_{k'},\mathcal{T}}\\
     &=\sum_{k'=1}^{K}
  \sum_{\substack{\mathcal{T}\in\mathfrak{T}:\\
                   k'\in\bar{\mathcal U}(\mathcal{T}),\ k\in\bar{\mathcal U}(\mathcal{T})}}
     \det \mathbf{H}[\mathcal{Q},\mathcal D_{k',\mathcal{T}}]\,\mathbf{w}_{d_{k'},\mathcal{T}}
 \label{eq:sv-residual-recipe} 
\end{align}
\end{subequations}

Next user U$_k$ decodes its desired subfiles based on $\mathbf{y}_{k,\mathcal{Q}}$ for all $\mathcal{Q}\in\binom{[q+t]}{q-1}$.  More precisely, for each set $\mathcal{S}=\{a_1,\ldots,a_q\}\in\binom{[q+t]}q$, assume that $a_1<\cdots<a_q$. User U$_k$ forms 
\begin{align}
\mathbf{y}_{k,\mathcal{S}}=\sum_{j=1}^{q}(-1)^{j-1}h_{a_j,k} \mathbf{y}_{k,\mathcal{S}\setminus\{a_j\}},
 \label{eq:sv-combiner}
\end{align}
where $h_{a_j,k}$ is the element located at the $a_j$-th row and $k$-th column of $\mathbf{H}$. All required labels $\mathcal{S}\setminus\{a_j\}$ occur among the broadcasts in~\eqref{eq:det-broadcast}, so these are local computations, not additional transmissions.

Recall that $[\mathbf{h}_k,\mathbf{H}[:,\mathcal{D}]]$ denotes horizontal concatenation with
$\mathbf{h}_k$ as the first column, followed by the columns indexed by $\mathcal{D}$. Then,
from~\eqref{eq:sv-combiner}, we have 
\begin{subequations}
 \begin{align}
\mathbf{y}_{k,\mathcal{S}} =&\sum_{j=1}^{q}(-1)^{j-1}h_{a_j,k}
                        \mathbf{y}_{k,\mathcal{S}\setminus\{a_j\}}\\
= &\sum_{i=1}^{q}(-1)^{i-1}h_{a_i,k}
   \left(
    \sum_{k'=1}^{K}
    \sum_{\substack{\mathcal{T}\in\mathfrak{T}:\\
            k'\in\bar{\mathcal U}(\mathcal{T}), k\in\bar{\mathcal U}(\mathcal{T})}}
    \det \mathbf{H}[\mathcal{S}\setminus\{a_i\},\mathcal D_{k',\mathcal{T}}]\,\mathbf{w}_{d_{k'},
    \mathcal{T}}
   \right)\\
= &\sum_{k'=1}^{K}
   \sum_{\substack{\mathcal{T}\in\mathfrak{T}:\\
           k'\in\bar{\mathcal U}(\mathcal{T}),\ k\in\bar{\mathcal U}(\mathcal{T})}}
   \left(
     \sum_{i=1}^{q}(-1)^{i-1}h_{a_i,k}
        \det \mathbf{H}[\mathcal{S}\setminus\{a_i\},\mathcal D_{k',\mathcal{T}}]
   \right)\mathbf{w}_{d_{k'},\mathcal{T}}\\
= &\sum_{k'=1}^{K}
   \sum_{\substack{\mathcal{T}\in\mathfrak{T}:\\
           k'\in\bar{\mathcal U}(\mathcal{T}),\ k\in\bar{\mathcal U}(\mathcal{T})}}
   \det\bigl([\mathbf{h}_k\ \mathbf{H}[:,\mathcal D_{k',\mathcal{T}}]\,][\mathcal{S},:]\bigr)
       \mathbf{w}_{d_{k'},\mathcal{T}}, \label{eq:last expansion}
 \end{align}
 \end{subequations}
 where $[\mathbf{h}_k,\ \mathbf{H}[:,\mathcal D_{k',\mathcal{T}}]\,][\mathcal{S},:]$ represents the sub-matrix of $[\mathbf{h}_k, \ \mathbf{H}[:,\mathcal D_{k',\mathcal{T}}]]$ including the rows with indices in $\mathcal{S}$. 

Consider a summand in~\eqref{eq:last expansion} with $k'\ne k$.
Its summation restrictions imply $k,k'\in\bar{\mathcal U}(\mathcal{T})$; thus $ k\in\bar{\mathcal U}(\mathcal{T})\setminus\{k'\}=\mathcal D_{k',\mathcal{T}}.$ 
Consequently, $\mathbf{H}[:,\mathcal D_{k',\mathcal{T}}]$ already contains the column
$\mathbf{h}_k$. Prepending another $\mathbf{h}_k$ produces two equal columns, which
remain equal after restricting the matrix to any row set $\mathcal{S}$.
A determinant with two equal columns is zero over every field, i.e.,
\[
 \det\bigl([\mathbf{h}_k, \mathbf{H}[:,\mathcal D_{k',\mathcal{T}}]\,][\mathcal{S},:]\bigr)=0,
 \qquad  k'\ne k,\quad k,k'\in\bar{\mathcal U}(\mathcal{T}).
\]
For $k'=k$, in contrast, $\mathcal D_{k,\mathcal{T}}=\bar{\mathcal U}(\mathcal{T})\setminus\{k\}$
does not contain $k$, so this repeated-column argument does not force the coefficient to vanish. As a result, we have
\begin{subequations}
 \begin{align}
\mathbf{y}_{k,\mathcal{S}} =&\sum_{\substack{\mathcal{T}\in\mathfrak{T}:\\ k\in\bar{\mathcal U}(\mathcal{T})}}
       \det\bigl([\mathbf{h}_k, \mathbf{H}[:,\mathcal D_{k,\mathcal{T}}]\,][\mathcal{S},:]\bigr)\mathbf{w}_{d_k,\mathcal{T}}+\sum_{\substack{k'\in[K]:\\ k'\ne k}}
       \sum_{\substack{\mathcal{T}\in\mathfrak{T}:\\ k,k'\in\bar{\mathcal U}(\mathcal{T})}}
       0\cdot \mathbf{w}_{d_{k'},\mathcal{T}}\\
= &\sum_{\mathcal{T}\in\mathcal E_k}
       \det\bigl([\mathbf{h}_k, \mathbf{H}[:,\mathcal D_{k,\mathcal{T}}]\,][\mathcal{S},:]\bigr)\mathbf{w}_{d_k,\mathcal{T}}.
 \end{align}
\end{subequations}
Naming the surviving determinant coefficient $C_k(\mathcal{S},\mathcal{T})$ gives
 \begin{align}
 &\mathbf{y}_{k,\mathcal{S}}
   =\sum_{\mathcal{T}\in\mathcal E_k} C_k(\mathcal{S},\mathcal{T})\,\mathbf{w}_{d_k,\mathcal{T}},
       \label{eq:det-desired-system}\\
 &C_k(\mathcal{S},\mathcal{T})
   :=\det\bigl([\mathbf{h}_k, \mathbf{H}[:,\mathcal D_{k,\mathcal{T}}]\,][\mathcal{S},:]\bigr).
 \end{align}

Each $q$-element row set $\mathcal{S}\in \binom{[q+t]}{q}$  supplies one equation in
\eqref{eq:det-desired-system}, giving $\binom{q+t}{q}$ rows.
Each missing support $\mathcal{T}\in\mathcal E_k$ supplies one unknown packet,
giving $J=\binom{q+t-1}{t}$ columns. The overall coefficient matrix is therefore
\[
 \mathbf{C}_k(\mathbf{H})
 :=\bigl[C_k(\mathcal{S},\mathcal{T})\bigr]_{
         \mathcal{S}\in\binom{[q+t]}q,\ \mathcal{T}\in\mathcal{E}_k}
 \in\mathbb{F}_p^{\binom{q+t}{q}\times\binom{q+t-1}{t}}.
\]
Let $\mathbf{y}_k$ list the values $\mathbf{y}_{k,\mathcal{S}}$ in that row ordering,
and let $\mathbf{w}^{\rm miss}_k$ list the unknown packets $\mathbf{w}_{d_k,\mathcal{T}}$ in
that column ordering. Stacking those
scalar equations in the form of~\eqref{eq:det-desired-system} consequently gives 
$\mathbf{y}_k=\mathbf{C}_k(\mathbf{H})\mathbf{w}^{\rm miss}_k$. 
If $\mathbf{C}_k(\mathbf{H})$ has column rank $J$, row rank equals column rank, so it
contains $J$ linearly independent rows. Let $\mathcal{R}_k$ denote the index set of these $J$ linearly independent rows. The square sub-matrix $\mathbf{C}_k(\mathbf{H})[\mathcal{R}_k,:]$ is then
nonsingular. Restricting the stacked system to those rows and
left-multiplying by its inverse gives an explicit recovery rule:
\[
 \begin{aligned}
 \mathbf{y}_k[\mathcal{R}_k]
   &=\mathbf{C}_k(\mathbf{H})[\mathcal{R}_k,:]\mathbf{w}^{\rm miss}_k,\\
 \mathbf{w}^{\rm miss}_k
   &=\bigl(\mathbf{C}_k(\mathbf{H})[\mathcal{R}_k,:]\bigr)^{-1}
        \mathbf{y}_k[\mathcal{R}_k].
 \end{aligned}
\]
Conversely, if the column rank is smaller than $J$, a nonzero null
vector exists, and adding it to a candidate unknown vector leaves
all these equations unchanged. Thus full column rank is exactly what
makes this derived linear system uniquely solvable. 

It remains to obtain one matrix $\mathbf{H}$, satisfying 
\begin{align}
    \mathbf{C}_k(\mathbf{H}) \ {\rm is \ full \ rank}, \ \forall k\in [K],
    \label{eq:key decoding constraint}
\end{align} 
which will be stated in the following lemma whose proof is included in Appendix~\ref{subsec:sv-proof}.
\begin{lemma}
    \label{lem: SZ proof}
    %With the selection of   $\mathbf{H}$ whose elements are i.i.d. over ${\mathbb{F}}_{p}$,~\eqref{eq:key decoding constraint} holds with high probability.
If the entries of $\mathbf{H}$ are chosen independently and uniformly
from $\mathbb{F}_p$, then~\eqref{eq:key decoding constraint} holds
with high probability when $p$ is sufficiently large.
\hfill $\square$
\end{lemma}
 By Lemma~\ref{lem: SZ proof}, it can be seen that a specific $\mathbf{H}$ satisfying~\eqref{eq:key decoding constraint} can be easily found. 
\begin{remark}
    \label{rem:IA dimension}
    In the proposed caching scheme, each transmitted message is a linear combination of the desired subfiles by all users.  Among all of these transmitted subfiles, 
    the number of subfiles which are not known or desired by each user (i.e., the number of interference subfiles) is $(q-1)\binom{q+t-1}{q-1}$. In the transmitted $\binom{q+t}{q-1}$ messages,   we align the interference of each user into a dimension at most $\binom{q+t-1}{q-2}$, while preserving $\binom{q+t-1}{q-1}$ desired-signal dimensions separable from interference.
\hfill $\square$
\end{remark}

\begin{remark}[Boundary cases] For $t=0$, all caches are empty; sending the requested files uses
at most $K$ file units. For $t\geq1$ and $q=0$, every packet is
locally known to every user, so no transmission is needed. For
$q=1$, we have  $|\mathfrak{T}|=K$ and each user misses exactly $q=1$
subfile. The single broadcast message 
\begin{equation}
 \mathbf{x}_{\emptyset}
 =\sum_{k'=1}^{K}\sum_{\mathcal{T}\in\mathfrak{T}:k'\in\bar{\mathcal{U}}(\mathcal{T})}
     \mathbf{w}_{d_{k'},\mathcal{T}}
 \label{eq:sv-q-one}
\end{equation}
suffices, since each user knows every summand except its own. This
is also~\eqref{eq:det-broadcast} with the determinant of the empty
matrix defined as one.   
\hfill $\square$
\end{remark} 
\subsection[Example with K=N=6, t=2, and L=2]{An Example with $K=N=6$, $t=2$, and $L=2$}
\label{subsec:sv-example}

$\bullet$ {\bf Placement phase:} For notational convenience, we use the index $i$ to denote both
user U$_i$ and cache node C$_i$. Accordingly, user $1$ refers to U$_1$, and cache $1$ refers to C$_1$.
We have $q=K-tL=2$ and $M=Nt/K=2$. The cyclically $2$-separated
supports of size two are
\begin{align}
 \mathfrak{T}=\{\{1,3\},\{1,4\},\{1,5\},\{2,4\},\{2,5\},\{2,6\}, 
               \{3,5\},\{3,6\},\{4,6\}\}.
 \label{eq:sv-example-supports}
\end{align}
Divide each file $\mathbf{w}_n$ where $n\in[6]$ into $F=9$ equal-length subfiles 
$(\mathbf{w}_{n,\mathcal{T}} : \mathcal{T}\in\mathfrak{T})$. Cache $c$ stores $\mathbf{w}_{n,\mathcal{T}}$ if and only if $c\in \mathcal{T}$, for every $n\in[6]$.

Recall that $\bar{\mathcal{U}}(\mathcal{T})$ denotes the users who cannot obtain  the subfiles $(\mathbf{w}_{n,\mathcal{T}} : n\in[N])$ from their connected caches.  Table~\ref{tab:sv-example} lists all nine supports and their two missing users. Consequently, each user misses three subfiles of its requested file and already knows the remaining six. For instance, user $1$ accesses cache $1,2$ and misses precisely the supports $\{3,5\},\{3,6\},\{4,6\}$.
\begin{table}[http!]
 \centering
 \caption{Cache supports and missing users in the example. A packet
 with support $\mathcal{T}$ is stored in both caches indexed by $\mathcal{T}$.}
 \label{tab:sv-example}
 \begin{tabular}{cc@{\qquad\qquad}cc@{\qquad\qquad}cc}
 \toprule
 $\mathcal{T}$ & $\bar{\mathcal{U}}(\mathcal{T})$ & $\mathcal{T}$ & $\bar{\mathcal{U}}(\mathcal{T})$ & $\mathcal{T}$ & $\bar{\mathcal{U}}(\mathcal{T})$\\
 \midrule
 $\{1,3\}$ & $\{4,5\}$ & $\{2,4\}$ & $\{5,6\}$ & $\{3,5\}$ & $\{1,6\}$\\
 $\{1,4\}$ & $\{2,5\}$ & $\{2,5\}$ & $\{3,6\}$ & $\{3,6\}$ & $\{1,4\}$\\
 $\{1,5\}$ & $\{2,3\}$ & $\{2,6\}$ & $\{3,4\}$ & $\{4,6\}$ & $\{1,2\}$\\
 \bottomrule
 \end{tabular}
\end{table}

$\bullet$ {\bf Delivery phase:} Assume that the demand vector is $\mathbf{d}=(1,2,3,4,5,6)$.
In this example, we can use the finite field $\mathbb{F}_{41}$. 
We then construct a generating matrix $\mathbf{H}\in {\mathbb{F}}^{4\times 6}_{41}$ whose elements are i.i.d. over $\mathbb{F}_{41}$; a possible selection could be 
\begin{equation}
 \mathbf{H}=[\mathbf{h}_1\ \mathbf{h}_2\ \mathbf{h}_3\ \mathbf{h}_4\ \mathbf{h}_5\ \mathbf{h}_6]
 =\begin{pmatrix}
 1&0&0&0&1&1\\
 0&1&0&0&1&2\\
 0&0&1&0&1&3\\
 0&0&0&1&1&4
 \end{pmatrix}.
 \label{eq:sv-example-H}
\end{equation}
Here $q=2$, so $q-1=1$ and the broadcast labels in~\eqref{eq:det-broadcast} are the four
singletons $\mathcal{Q}=\{r\}$, $r\in[4]$. In addition, for each $\mathcal{T}\in \mathfrak{T}$, $\bar{\mathcal U}(\mathcal{T})$ only contains two elements. 
Hence, the broadcast messages in~\eqref{eq:det-broadcast} are
\begin{equation}
 \mathbf{x}_{\{r\}}
 =\sum_{\substack{\mathcal{T}\in\mathfrak{T}\\
                   \bar{\mathcal{U}}(\mathcal{T})=\{a,b\},\ a<b}}
       \bigl(\det[h_{r,b}]\mathbf{w}_{a,\mathcal{T}}+\det[h_{r,a}]\mathbf{w}_{b,\mathcal{T}}\bigr),
 \quad \forall r\in[4].
 \label{eq:sv-example-specialized-broadcast}
\end{equation}
For instance, when $\mathcal{T}=\{3,5\}$, we have $\bar{\mathcal U}(\{3,5\})=\{1,6\}$. Thus, the coefficients of $\mathbf{w}_{1,\{3,5\}}$ and $\mathbf{w}_{6,\{3,5\}}$ in $\mathbf{x}_{\{r\}}$ are
$\det \mathbf{H}[\{r\},\{6\}]$ and $\det \mathbf{H}[\{r\},\{1\}]$, respectively.
Using~\eqref{eq:sv-example-H}, these coefficient pairs are
$(1,1)$, $(2,0)$, $(3,0)$, and $(4,0)$ for $r=1,2,3,4$.

Evaluating every determinant in \eqref{eq:sv-example-specialized-broadcast} and omitting zero terms gives the following four transmitted linear combinations:
\begin{subequations}
 \begin{align}
 \mathbf{x}_{\{1\}}={}&\mathbf{w}_{6,\{3,5\}}+\mathbf{w}_{4,\{3,6\}}+\mathbf{w}_{2,\{4,6\}}+\mathbf{w}_{4,\{1,3\}}+\mathbf{w}_{2,\{1,4\}}+\mathbf{w}_{6,\{2,4\}}\notag\\ &+\mathbf{w}_{5,\{2,4\}}+\mathbf{w}_{3,\{2,5\}}+\mathbf{w}_{1,\{3,5\}},
 \label{eq:sv-example-X1}\\
 \mathbf{x}_{\{2\}}={}&\mathbf{w}_{5,\{1,4\}}+\mathbf{w}_{3,\{1,5\}}+\mathbf{w}_{1,\{4,6\}}+\mathbf{w}_{4,\{1,3\}}+\mathbf{w}_{2,\{1,4\}}+\mathbf{w}_{6,\{2,4\}}\notag\\ &+2\mathbf{w}_{5,\{2,4\}}+2\mathbf{w}_{3,\{2,5\}}+2\mathbf{w}_{1,\{3,5\}},
 \label{eq:sv-example-X2}\\
 \mathbf{x}_{\{3\}}={}&\mathbf{w}_{2,\{1,5\}}+\mathbf{w}_{6,\{2,5\}}+\mathbf{w}_{4,\{2,6\}}+\mathbf{w}_{4,\{1,3\}}+\mathbf{w}_{2,\{1,4\}}+\mathbf{w}_{6,\{2,4\}}\notag\\ &+3\mathbf{w}_{5,\{2,4\}}+3\mathbf{w}_{3,\{2,5\}}+3\mathbf{w}_{1,\{3,5\}},
 \label{eq:sv-example-X3}\\
 \mathbf{x}_{\{4\}}={}&\mathbf{w}_{5,\{1,3\}}+\mathbf{w}_{3,\{2,6\}}+\mathbf{w}_{1,\{3,6\}}+\mathbf{w}_{4,\{1,3\}}+\mathbf{w}_{2,\{1,4\}}+\mathbf{w}_{6,\{2,4\}}\notag\\ &+4\mathbf{w}_{5,\{2,4\}}+4\mathbf{w}_{3,\{2,5\}}+4\mathbf{w}_{1,\{3,5\}}.
 \label{eq:sv-example-X4}
  \end{align}
\end{subequations}
Recall that for each $k\in [K]$,  $\mathbf{y}_{k,\mathcal{Q}}$ denotes the residual of broadcast $\mathbf{x}_\mathcal{Q}$ after user $k$
removes all the subfiles which are stored in its connected caches. 

We first consider user~$1$, which is connected to caches $1$ and $2$. User $1$ demands $\mathbf{w}_1$ and can retrieve all the subfiles of $\mathbf{w}_1$ from caches $1$ and $2$, except $\mathbf{w}_{1,\{3,5\}}, \mathbf{w}_{1,\{4,6\}},\allowbreak\mathbf{w}_{1,\{3,6\}}$. Hence,   user $1$ needs to recover these $3$ subfiles from the delivery phase.  

By receiving $\mathbf{x}_{\{1\}}$ and using the contents cached by caches $1$ and $2$, 
user $1$ can
compute
\begin{align*}
    \mathbf{y}_{1,\{1\}}&=\mathbf{x}_{\{1\}}
 -\mathbf{w}_{4,\{1,3\}}-\mathbf{w}_{2,\{1,4\}}-\mathbf{w}_{6,\{2,4\}}
 -\mathbf{w}_{3,\{2,5\}}-\mathbf{w}_{5,\{2,4\}}\\
    &=\mathbf{w}_{2,\{4,6\}}+\mathbf{w}_{4,\{3,6\}}+\mathbf{w}_{6,\{3,5\}}+\mathbf{w}_{1,\{3,5\}},
\end{align*}
because all the subtracted packets are available in caches $1$ and $2$ connecting to user $1$.
Applying
the same subtraction to each broadcast message, user~$1$ obtains
\begin{subequations}
 \begin{align}
 \mathbf{y}_{1,\{1\}}&=\mathbf{w}_{2,\{4,6\}}+\mathbf{w}_{4,\{3,6\}}+\mathbf{w}_{6,\{3,5\}}+\mathbf{w}_{1,\{3,5\}},\\
 \mathbf{y}_{1,\{2\}}&=\mathbf{w}_{1,\{4,6\}}+2\mathbf{w}_{1,\{3,5\}}, \label{eq:Y12}\\
 \mathbf{y}_{1,\{3\}}&=3\mathbf{w}_{1,\{3,5\}},\label{eq:Y13}\\
 \mathbf{y}_{1,\{4\}}&=\mathbf{w}_{1,\{3,6\}}+4\mathbf{w}_{1,\{3,5\}}.\label{eq:Y14}
 \end{align}
 \label{eq:sv-example-user1-equations}
Then user $1$ can recover the required subfiles $\mathbf{w}_{1,\{3,5\}}, \mathbf{w}_{1,\{4,6\}},\allowbreak\mathbf{w}_{1,\{3,6\}}$ from~\eqref{eq:Y12}-\eqref{eq:Y14}. Similarly, users $2,3,4$ can recover their required subfiles.
\end{subequations}

We then focus on user $5$, which is connected to caches $5$ and $6$. 
User $5$ demands $\mathbf{w}_5$ and can retrieve all the subfiles of $\mathbf{w}_5$ from caches $5$ and $6$, except $\mathbf{w}_{5,\{1,3\}},\allowbreak\mathbf{w}_{5,\{1,4\}},\allowbreak\mathbf{w}_{5,\{2,4\}}$. Hence,   user $5$ needs to recover these $3$ subfiles from the delivery phase. By receiving $\mathbf{x}_{\{1\}}, \mathbf{x}_{\{2\}}, \mathbf{x}_{\{3\}}, \mathbf{x}_{\{4\}}$ and using the  content stored in caches $5$ and $6$, 
user $5$ can compute 
\begin{subequations}
 \begin{align}
 \mathbf{y}_{5,\{1\}}&=\mathbf{w}_{4,\{1,3\}}+\mathbf{w}_{2,\{1,4\}}+\mathbf{w}_{6,\{2,4\}}+\mathbf{w}_{5,\{2,4\}},\\
 \mathbf{y}_{5,\{2\}}&=\mathbf{w}_{4,\{1,3\}}+\mathbf{w}_{2,\{1,4\}}+\mathbf{w}_{6,\{2,4\}}+\mathbf{w}_{5,\{1,4\}}+2\mathbf{w}_{5,\{2,4\}},\\
 \mathbf{y}_{5,\{3\}}&=\mathbf{w}_{4,\{1,3\}}+\mathbf{w}_{2,\{1,4\}}+\mathbf{w}_{6,\{2,4\}}+3\mathbf{w}_{5,\{2,4\}},\\
 \mathbf{y}_{5,\{4\}}&=\mathbf{w}_{4,\{1,3\}}+\mathbf{w}_{2,\{1,4\}}+\mathbf{w}_{6,\{2,4\}}+\mathbf{w}_{5,\{1,3\}}+4\mathbf{w}_{5,\{2,4\}}.
 \end{align}
 \label{eq:sv-example-user5-equations}
\end{subequations}
Note that  $\mathbf{w}_{4,\{1,3\}}, \mathbf{w}_{2,\{1,4\}}, \mathbf{w}_{6,\{2,4\}}$ are the interferences to user $5$; it can be seen that the interferences are aligned into the same one-dimensional subspace in $\mathbf{y}_{5,\{1\}}, \mathbf{y}_{5,\{2\}}, \mathbf{y}_{5,\{3\}}, \mathbf{y}_{5,\{4\}}$.
 Consequently, user $5$ can further  compute 
\begin{subequations}
 \begin{align}
 \mathbf{y}_{5,\{1,2\}}
 &=\mathbf{y}_{5,\{2\}}-\mathbf{y}_{5,\{1\}}
 =\mathbf{w}_{5,\{1,4\}}+\mathbf{w}_{5,\{2,4\}},\\
 \mathbf{y}_{5,\{1,3\}}
 &=\mathbf{y}_{5,\{3\}}-\mathbf{y}_{5,\{1\}}
 =2\mathbf{w}_{5,\{2,4\}},\\
 \mathbf{y}_{5,\{1,4\}}
 &=\mathbf{y}_{5,\{4\}}-\mathbf{y}_{5,\{1\}}
 =\mathbf{w}_{5,\{1,3\}}+3\mathbf{w}_{5,\{2,4\}}.
 \end{align}
 \label{eq:sv-example-user5-combinations}
\end{subequations}
The aggregate interferences have been canceled in each equation; thus 
user $5$ can recover $\mathbf{w}_{5,\{1,3\}},\allowbreak\mathbf{w}_{5,\{1,4\}},\allowbreak\mathbf{w}_{5,\{2,4\}}$. 
Similarly, user $6$ can also recover its required subfiles. 

The achieved load is
$
 R=\frac49=\frac{q(q+t)}{K(t+1)},
$
which matches the fixed-placement converse of Theorem~\ref{thm:main theorem}. The observations
identify the two properties needed in general: unknown interference
must vanish after combining broadcasts, and the remaining desired
coefficient matrix must have full column rank.

\begin{remark}[Comparison with existing schemes]
For the same cyclic network with $K=N=6$, $L=2$, and per-cache
memory $M=2$ (i.e., $M/N=1/3$ and $t=KM/N=2$), we compare the proposed scheme with the existing schemes: The HKD scheme with MAN delivery in~\cite{HKD}, the  RK1 scheme in~\cite{RK}, the CWLZC scheme in~\cite{CWLZC} achieve a load of $2/3$; the SPE scheme in~\cite{SPE} and the KK scheme~\cite{KK2023} achieve a load of $1/2$. It can be seen that in this example, the proposed scheme is strictly better than the above schemes. However, the RK2 scheme (which also uses the same cyclically $L$-separated placement) already attains the same load $4/9$.  However, as illustrated in Section~\ref{sub:theoretic performance}, the proposed scheme is strictly better than RK2; for example, when $K=N=7, M=L=2$, the proposed scheme achieves the load $5/7$, while RK2 achieves $11/14$.
\hfill $\square$
\end{remark}

\section{Proof of Theorem~\ref{thm:uncoded converse}: Global Uncoded-Placement Converse}
\label{sec:global-converse}
Under the constraint of uncoded cache placement, we can divide each file $\mathbf{w}_n$ where $n\in [N]$ into subfiles, $\mathbf{w}_n=\{\mathbf{w}_{n,\mathcal{T}}:\mathcal{T}\subseteq[K]\}$,  where $\mathbf{w}_{n,\mathcal{T}}$ contains the bits of $\mathbf{w}_n$ stored in exactly the caches indexed by $\mathcal{T}$.  The file-size and individual cache-memory constraints imply that
\begin{align}
 &\sum_{\mathcal{T}\subseteq[K]}|\mathbf{w}_{n,\mathcal{T}}|=B,
 \ n\in[N],
 \label{eq:t2-file-constraint}\\
& \sum_{n=1}^{N}\sum_{\mathcal{T}\subseteq[K]: k'\in \mathcal{T}}
 |\mathbf{w}_{n,\mathcal{T}}|\leq MB,
 \ k'\in[K].
 \label{eq:t2-cache-constraint}
\end{align}
%\noindent\textbf{2) Acyclic sets associated with cyclic user orders.}
Since $N\geq K$, we can choose the set of distinct-demand vectors
\begin{align}
 \mathcal{D}_{\mathrm{dist}}
 =\{\mathbf{d}\in[N]^K:d_k\neq d_v\text{ for all }k\neq v\},
 \label{eq:t2-distinct-demands}
\end{align}
where in each distinct-demand vector, each user requests a different file.

For each starting user U$_k$, $k\in[K]$, consider the cyclic
order $\pi_k(k_1)=\cyc{k+k_1-1}$, where $k_1\in[K]$.
Note that $\bigcup_{k_2=1}^{k_1}\mathcal{A}_{\pi_k(k_2)}$ represents
the union of the indices of cache nodes accessible to the first
$k_1$ users in this cyclic order, where $\mathcal{A}_{k'}$, defined
in~\eqref{eq:access}, denotes the set of indices of cache nodes
accessible to user U$_{k'}$. We retain the following subfiles
at level $k_1$:
\begin{equation}
 \mathcal{V}_{k,k_1}(\mathbf{d})
 =\left\{
   \mathbf{w}_{d_{\pi_k(k_1)},\mathcal{T}}:
   \mathcal{T}\subseteq[K],\;
   \mathcal{T}\cap
   \bigcup_{k_2=1}^{k_1}\mathcal{A}_{\pi_k(k_2)}=\emptyset
  \right\}.
 \label{eq:t2-selected-level}
\end{equation}
Every retained subfile at level $k_1$ is requested by user
U$_{\pi_k(k_1)}$ and is unavailable from the cache nodes accessible
to any of the users U$_{\pi_k(1)},\ldots,$ U$_{\pi_k(k_1)}$.
Hence, the sub-graph induced by the nodes in
$\bigcup_{k_1=1}^{K}\mathcal{V}_{k,k_1}(\mathbf{d})$ is acyclic.
The acyclic index-coding bound~\cite{ZYST,WTP2020} yields
\begin{equation}
 R B\geq H(\mathbf{X}_{\mathbf{d}})
 \geq\sum_{k_1=1}^{K}
       \sum_{\substack{
         \mathcal{T}\subseteq[K]:\\
         \mathcal{T}\cap
         \bigcup_{k_2=1}^{k_1}\mathcal{A}_{\pi_k(k_2)}=\emptyset}}
       |\mathbf{w}_{d_{\pi_k(k_1)},\mathcal{T}}|.
 \label{eq:t2-acyclic-bound}
\end{equation}
For each $\mathcal{T}\subseteq[K]$, define
$x_{\mathcal{T}}:=\frac{1}{NB}\sum_{n=1}^{N}|\mathbf{w}_{n,\mathcal{T}}|$.
Considering all distinct-demand vectors and averaging together
all the resulting inequalities as in~\eqref{eq:t2-acyclic-bound},
we can obtain
\begin{equation}
 R\geq\sum_{k_1=1}^{K} \sum_{\substack{ \mathcal{T}\subseteq[K]:\\
         \mathcal{T}\cap \bigcup_{k_2=1}^{k_1}\mathcal{A}_{\pi_k(k_2)}=\emptyset}}
       x_{\mathcal{T}}.
 \label{eq:t2-demand-averaged-bound}
\end{equation}
Next, we define
\[
 C_{k_1}(\mathcal{T}) =\left| \left\{ k\in[K]: \mathcal{T}\cap \bigcup_{k_2=1}^{k_1}\mathcal{A}_{\pi_k(k_2)}=\emptyset
   \right\}
  \right|.
\]
Averaging~\eqref{eq:t2-demand-averaged-bound} over all $K$ starting
users gives
\begin{equation}
 \begin{aligned}
 R\geq \frac{1}{K}\sum_{k=1}^{K}\sum_{k_1=1}^{K}
       \sum_{
         \mathcal{T}\subseteq[K],
         \mathcal{T}\cap         \bigcup_{k_2=1}^{k_1}\mathcal{A}_{\pi_k(k_2)}=\emptyset}       x_{\mathcal{T}}=\sum_{\mathcal{T}\subseteq[K]}x_{\mathcal{T}}
   \left(\frac{1}{K}\sum_{k_1=1}^{K}C_{k_1}(\mathcal{T})\right).
 \end{aligned}
 \label{eq:t2-cyclic-averaged-bound}
\end{equation}
For each $k_1\in[K]$, the union $\bigcup_{k_2=1}^{k_1}\mathcal{A}_{\pi_k(k_2)}$ has length $\min\{K,L+k_1-1\}$. So as $k$ ranges over $[K]$, each fixed cache node C$_{\lambda}$, $\lambda\in[K]$, belongs to exactly $\min\{K,L+k_1-1\}$ such
intervals, i.e.,
\begin{equation}
 \left|
  \left\{
   k\in[K]:
   \lambda\in\bigcup_{k_2=1}^{k_1}\mathcal{A}_{\pi_k(k_2)}
  \right\}
 \right|
 =\min\{K,L+k_1-1\},
 \qquad \forall \lambda\in[K].
 \label{eq:t2-incidence-count}
\end{equation}
For any support $\mathcal{T}$ with $|\mathcal{T}|=r$, the set of
starting users whose interval intersects $\mathcal{T}$ is a union
over the cache nodes indexed by $\mathcal{T}$. Consequently,
\begin{subequations}
\begin{align}
 &\left|
   \left\{
    k\in[K]:
    \mathcal{T}\cap
    \bigcup_{k_2=1}^{k_1}\mathcal{A}_{\pi_k(k_2)}\neq\emptyset
   \right\}
  \right| \notag\\
 =&
  \left|
   \bigcup_{\lambda\in\mathcal{T}}
   \left\{
    k\in[K]:
    \lambda\in\bigcup_{k_2=1}^{k_1}\mathcal{A}_{\pi_k(k_2)}
   \right\}
  \right|
 \label{eq:transfer}\\
 \leq&
  \min\left\{
   K,\sum_{\lambda\in\mathcal{T}}
   \left|
    \left\{
     k\in[K]:
     \lambda\in\bigcup_{k_2=1}^{k_1}\mathcal{A}_{\pi_k(k_2)}
    \right\}
   \right|
  \right\}\\
 =&\min\{K,r(L+k_1-1)\},
 \label{eq:t2-union-count}
\end{align}\end{subequations}
where~\eqref{eq:transfer} follows since
$\mathcal{T}\cap
 \bigcup_{k_2=1}^{k_1}\mathcal{A}_{\pi_k(k_2)}\neq\emptyset$
if and only if there exists $\lambda\in\mathcal{T}$ such that
$\lambda\in\bigcup_{k_2=1}^{k_1}\mathcal{A}_{\pi_k(k_2)}$,
and~\eqref{eq:t2-union-count} comes
from~\eqref{eq:t2-incidence-count}.

Taking the complement among the $K$ possible starting users,
\begin{subequations}
\begin{align}
 C_{k_1}(\mathcal{T})
 &=K-
   \left|
    \left\{
     k\in[K]:
     \mathcal{T}\cap
     \bigcup_{k_2=1}^{k_1}\mathcal{A}_{\pi_k(k_2)}\neq\emptyset
    \right\}
   \right|\\
 &\geq K-\min\{K,r(L+k_1-1)\}
  =\pos{K-r(L+k_1-1)}.
 \label{eq:t2-avoidance-bound}
\end{align}
\end{subequations}
The last equality in~\eqref{eq:t2-avoidance-bound} follows from $K-\min\{K,a\}=(K-a)^+$ for $a\geq0$, with $a=r(L+k_1-1)$.
Summing~\eqref{eq:t2-avoidance-bound} over $k_1\in[K]$ gives
\begin{align}
 \frac{1}{K}\sum_{k_1=1}^{K}C_{k_1}(\mathcal{T})
 \geq\frac{1}{K}\sum_{k_1=1}^{K}\pos{K-r(L+k_1-1)}
 \notag=\frac{1}{K}\sum_{k_1=0}^{K-1}\pos{K-r(L+k_1)}
  =\Phi_r.
 \label{eq:t2-coefficient-bound}
\end{align}
Let $x_r$ be the fraction of library bits stored in exactly $r$
cache nodes:
\begin{equation}
 \begin{aligned}
 x_r
 &=\sum_{\mathcal{T}\in\binom{[K]}{r}}x_{\mathcal{T}}=\frac{1}{NB}\sum_{n=1}^{N}
   \sum_{\mathcal{T}\in\binom{[K]}{r}}
   |\mathbf{w}_{n,\mathcal{T}}|,
 \qquad \forall r\in[0:K].
 \end{aligned}
 \label{eq:t2-replication-fractions}
\end{equation}
Substituting~\eqref{eq:t2-avoidance-bound} into~\eqref{eq:t2-cyclic-averaged-bound} yields
\begin{equation}
 R\geq\sum_{r=0}^{K}\Phi_r x_r.
 \label{eq:t2-weighted-bound}
\end{equation}
Note that for each integer $r\geq\lceil K/L\rceil$, one has
$rL\geq K$ and thus $\Phi_r=0$.

The file-size constraints in~\eqref{eq:t2-file-constraint}
imply $x_r\geq0$ and
\begin{equation}
 \sum_{r=0}^{K}x_r=1.
 \label{eq:t2-fraction-constraints}
\end{equation}
Moreover, summing~\eqref{eq:t2-cache-constraint} over all cache nodes
counts each bit of $\mathbf{w}_{n,\mathcal{T}}$ exactly
$|\mathcal{T}|$ times. Hence,
\begin{align}
 \sum_{r=0}^{K}r x_r
 =\frac{1}{NB}\sum_{n=1}^{N}
   \sum_{\mathcal{T}\subseteq[K]}
   |\mathcal{T}|\,|\mathbf{w}_{n,\mathcal{T}}|
=\frac{1}{NB}\sum_{\lambda=1}^{K}\sum_{n=1}^{N}
   \sum_{  \mathcal{T}\subseteq[K],
     \lambda\in\mathcal{T}}
   |\mathbf{w}_{n,\mathcal{T}}|\leq\frac{KM}{N}.
 \label{eq:t2-memory-moment}
\end{align}

As in~\cite{YMA2018}, we can lower bound the right-hand side
of~\eqref{eq:t2-weighted-bound} by using Jensen's inequality
and the monotonicity of ${\rm Conv}(\Phi_r)$
(i.e., the convex lower envelope of $\Phi_r$ in terms of $r$).
Together with~\eqref{eq:t2-memory-moment}, this gives
\begin{align*}
 R^{\star}_{\rm u}(M)
 \geq
 \left[
  {\rm Conv}
  \left(
   \left\{
    \left(\frac{Nr}{K},\Phi_r\right):
    r\in\left[0:\left\lceil\frac{K}{L}\right\rceil\right]
   \right\}
  \right)
 \right](M),
 \qquad
 0\leq M\leq\frac{N}{K}\left\lceil\frac{K}{L}\right\rceil.
\end{align*}
For $\frac{N}{K}\lceil K/L\rceil\leq M\leq N$, the lower bound
is zero. This coincides with Theorem~\ref{thm:uncoded converse}.

\section{Conclusion}
\label{sec:conclusion}

In this paper, we investigated the MACC problem under the cyclically $L$-separated placement. By refining the acyclic index coding analysis and accounting for all valid cache supports, we derived a converse bound that improves upon an existing bound for the same placement. We then proposed a linear coded caching scheme based on a novel determinant-based interference alignment strategy, which simultaneously aligns interference across all users while preserving the decodability of their desired signals. 
The proposed scheme exactly matches the derived converse bound, thereby characterizing the optimal transmission load under the cyclically $L$-separated placement.
Furthermore, we established a general converse bound applicable to arbitrary uncoded cache placements. 
By combining the proposed scheme with memory sharing, we proved that the resulting scheme is order optimal within a factor of $2$ under uncoded cache placement for $N\geq K$.
Theoretical and numerical comparisons demonstrated that our scheme establishes the exact optimal load under the cyclically $L$-separated placement and the same subpacketization in certain parameter regimes. Moreover, compared to some existing schemes with linear subpacketization, our scheme achieved the same level of subpacketization and lower transmission load for some system parameters. Ongoing work includes the extension of the proposed scheme to more general networks, combination networks, and hierarchical networks.

\paragraph*{Disclosure of AI Usage}
 After obtaining the main idea of the determinant-based MACC scheme with some concrete construction examples, GPT-6 was used to help the authors write down the complete description of the scheme and its decodability proof. The authors also used GPT-6 to proofread the manuscript and provide an initial theoretical comparison to the existing schemes. The authors have   manually verified and significantly simplified  the  AI-assisted proofs, and also manually  revised the writing of the manuscript. The authors take full responsibility for the correctness of the results.

\appendices

\section{Proof of Theorem~\ref{thm:uncoded-two-approx}} \label{app:uncoded-two-approx}
According to Theorem~\ref{thm:uncoded converse}, the lower bound on the optimal transmission load under uncoded placement is given by the lower convex envelope of the set of points $\{(Nt/K,\Phi_t):t \in [0:\lceil K/L \rceil]\}$. In order to establish the factor-two approximation guarantee, it suffices to show that for each integer $t \in [0,\lceil K/L\rceil]$, there exists an uncoded placement scheme that achieves a transmission load $ R_t \leq 2\Phi_t$ at the memory point $M = Nt/K$. This result can then be extended to arbitrary cache sizes through memory sharing.

We first consider an integer $1\leq t< \lceil K/L\rceil$ and define $q=K-tL>0$. The achievable transmission load in Theorem~\ref{thm:main theorem} is given by 
\begin{equation}
    R_t=\frac{q(q+t)}{K(t+1)}.
    \label {eq:app-achievable-corner}
\end{equation}
Write $q=at+b$, where the quotient is $a=\lfloor q/t\rfloor$ and the remainder satisfies $0\leq b<t$. The coefficient satisfies 
\begin{equation*}
    K\Phi_t=\sum_{i=0}^{K-1}(q-ti)^+=\sum_{i=0}^{a}(q-ti)=\frac{ta(a+1)}{2}+(a+1)b=\frac{q(q+t)+b(t-b)}{2t},
\end{equation*}
where $(x)^+=\max\{0,x\}$. Since $0\le b<t$, we have $b(t-b)\geq 0$, hence, $K\Phi_t\geq \frac{q(q+t)}{2t}$. By \eqref{eq:app-achievable-corner}, we have $R_t\leq \frac{2t}{t+1}\Phi_t$. For any positive integer $t \geq 1$, $\frac{2t}{t+1}\leq 2$ holds. Thus 
 \begin{equation}
     R_t \leq \frac{2t}{t+1}\Phi_t \leq 2\Phi_t.
 \end{equation}
 Next, we verify the two endpoints of the interval.
 \begin{itemize}
     \item When $t=0$, we have $M=0$, i.e., all cache nodes have zero memory; all files must be transmitted via broadcast. At this point, the transmission load $R_0=K=\Phi_0$.
     \item When $t=\lceil K/L\rceil$, we have the converse coefficient $\Phi_t=0$. We need to show that, for a cache capacity $M=\frac{N}{K}\lceil\frac{K}{L}\rceil$, there exists a coded caching scheme where the transmission load is zero under the uncoded placement. 

     Define $\mathcal{C}\triangleq \{1,1+L,\ldots,1+(\lceil K/L\rceil-1)L\}.$ For each $s\in[0:K-1]$, define the cyclic shift of $\mathcal{C}$ by $$\mathcal{C}_s\triangleq\left\{\langle c+s\rangle_K : c\in\mathcal{C}\right\}.$$ Split each file $\mathbf{w}_n$ into $K$ equal-size subfiles $\mathbf{w}_n=\bigl(\mathbf{w}_{n,s}:s\in[0:K-1]\bigr)$, and store $\mathbf{w}_{n,s}$ in cache node C$_k$ if and only if $k\in\mathcal{C}_s$.
     Since every cyclic gap in $\mathcal{C}_s$ is at most $L$, the $L$ consecutive cache nodes accessible to each user intersect $\mathcal{C}_s$ for every $s\in[0:K-1]$. Hence, every user can retrieve all subfiles of its requested file from its accessible caches, and the transmission load is zero. Moreover, each cache node belongs to exactly $\lceil\frac{K}{L}\rceil$ shifted supports and therefore stores an $\lceil\frac{K}{L}\rceil/K$ fraction of each file. Thus, the required cache memory is $\frac{N}{K}\lceil\frac{K}{L}\rceil$.

 \end{itemize}
 Now fix any $M\in\left[0,\frac{N}{K}\left\lceil\frac{K}{L}\right\rceil\right]$, and let $R_{\mathrm{lb}}$  and $ R_{\mathrm{u}}^\star$ denote the value of the lower convex envelope in Theorem~\ref{thm:uncoded converse} and the optimal transmission load
under uncoded placement at this memory size, respectively. Choose nonnegative weights $\lambda_0,\ldots,\lambda_{\lceil K/L \rceil}$ attaining this envelope:
\begin{equation}
\sum_{t=0}^{\lceil K/L\rceil}\lambda_t=1,\ 
\sum_{t=0}^{\lceil K/L\rceil}\lambda_t Nt/K =M,\ 
\sum_{t=0}^{\lceil K/L\rceil}\lambda_t\Phi_t=R_{\mathrm{lb}}.
\label{eq:app-envelope-weights}
\end{equation}
Memory sharing among the achievable corner points using these weights
yields an achievable transmission load
\begin{align}
R_{\mathrm{ms}} = \sum_{t=0}^{\lceil K/L\rceil}\lambda_t R_t  \leq 2\sum_{t=0}^{\lceil K/L\rceil}\lambda_t\Phi_t
=2R_{\mathrm{lb}} \leq 2R_{\mathrm{u}}^\star.
\label{eq:app-full-region-factor-two}
\end{align} 

Since this memory-sharing construction uses only uncoded placements, we also have $R_{\mathrm{u}} ^\star \leq R_{\mathrm{ms}}$. For $\frac{N}{K}\left\lceil\frac{K}{L}\right\rceil \leq M \leq N$, the zero-load placement above remains feasible. Therefore, throughout $M\in[0,N]$,
\begin{equation}
    R_{\mathrm{u}}^\star \leq R_{\mathrm{ms}} \leq 2R_{\mathrm{u}}^\star.
\end{equation}

\section{Proof of Corollary~\ref{cor:optimal regime}}
\label{sec:proof of cor}
When $KM/N\in[0,1]$, we will show that the achievable load of
Theorem~\ref{thm:main theorem} and the converse of Theorem~\ref{thm:uncoded converse} coincide throughout this interval.

We first consider the achievable load. 
At $M=0$, transmitting the requested files without caching achieves
worst-case load $R_0=K$, since $N\geq K$.
At $M=N/K$, substituting
$t=1$ and $q=K-L$ into the achievable load in Theorem~\ref{thm:main theorem} gives
\[
 \left.\frac{q(q+t)}{K(t+1)}\right|_{t=1,\,q=K-L}
     =\frac{(K-L)(K-L+1)}{2K}.
\]
For a general $M\in[0,N/K]$, by memory sharing between $(0,K)$ and $(N/K, \frac{(K-L)(K-L+1)}{2K})$, the achieved load is 
\begin{align}
    K-\frac{KM}{N}\left[K-\frac{(K-L)(K-L+1)}{2K}\right].
    \label{eq:first-segment-achievability}
\end{align}

We now consider the converse bound. Recall the integer-point coefficients in Theorem~\ref{thm:uncoded converse},
\[
 \Phi_t=\frac{1}{K}\sum_{i=0}^{K-1}
          \bigl(K-t(L+i)\bigr)^+,
\]
whose first two values are
\begin{align*}
 \Phi_0&=\frac{1}{K}\sum_{i=0}^{K-1}K=K,\\
 \Phi_1&=\frac{1}{K}\sum_{i=0}^{K-1}(K-L-i)^+
        =\frac{1}{K}\sum_{j=1}^{K-L}j
        =\frac{(K-L)(K-L+1)}{2K}.
\end{align*}
We next prove explicitly that the lower
convex envelope used in Theorem~\ref{thm:uncoded converse} coincides with the line segment
joining $(0,\Phi_0)$ and $(1,\Phi_1)$ throughout $0\leq KM/N\leq1$;
in particular, no convex combination involving later integer points
can lower this segment.

Note that the piecewise-linear
interpolant through ordered samples of a convex function coincides with
the lower convex envelope of those samples. Here
\[
 f(z)=\frac{1}{K}\sum_{i=0}^{K-1}
             \max\{K-z(L+i),0\}, \ \textrm{where } z\geq0,
\]
is convex as a sum of maxima of affine functions, and $f(t)=\Phi_t$.
In addition, one can check that its integer samples have non-decreasing successive slopes
$\Phi_{t+1}-\Phi_t$ with respect to $t$; thus their piecewise-linear interpolant is convex
and is precisely their lower convex envelope. Denoting this envelope on $[0,\lceil K/L\rceil]$ by $\underline\Phi$, we obtain that when $0\leq KM/N \leq 1$, 
\begin{equation}
 \underline\Phi(KM/N)
 =(1-\frac{KM}{N})\Phi_0+ \frac{KM}{N}\Phi_1,
 \qquad 0\leq KM/N\leq1.
 \label{eq:first-segment-hull-equality}
\end{equation}
Finally, Theorem~\ref{thm:uncoded converse} gives $R^{\star}_{\rm u}(M)\geq\underline\Phi(KM/N)$.
Substituting~\eqref{eq:first-segment-hull-equality} and the values of
$\Phi_0$ and $\Phi_1$, we obtain
\begin{equation}
 R^{\star}_{\rm u}(M)\geq(1-KM/N)\Phi_0+KM/N\Phi_1
 =K-\frac{KM}{N}\left[K-\frac{(K-L)(K-L+1)}{2K}\right],
 \label{eq:first-segment-converse}
\end{equation}
 which coincides with~\eqref{eq:first-segment-achievability}.

\section{Proof of Lemma~\ref{lem: SZ proof}}
\label{subsec:sv-proof}
Throughout this subsection, assume $q\geq2$, where the boundary cases were discussed at the end of  Section~\ref{sub:general encoding rule}. We first reduce the full-rank requirement to the non-zero of a determinant polynomial and apply the Schwartz--Zippel lemma. We then prove the required non-zero lemma by constructing a specific auxiliary realization of the generating matrix.

Fix a user U$_k\in[K]$. Recall that $\mathcal{E}_k=\{\mathcal{T}\in\mathfrak{T}:k\in\bar{\mathcal{U}}(\mathcal{T})\}$, $\mathcal{D}_{k,\mathcal{T}}=\bar{\mathcal{U}}(\mathcal{T})\setminus\{k\}$, $|\mathcal{E}_k|=\binom{q+t-1}{q-1}$, and $|\mathcal{D}_{k,\mathcal{T}}|=q-1$. The interference cancellation already established there gives
\begin{subequations}
 \begin{align}
  \mathbf{y}_k&=\mathbf{C}_k(\mathbf{H})\mathbf{w}^{\rm miss}_k,\\
  \mathbf{C}_k(\mathbf{H})&=\bigl[C_k(\mathcal{S},\mathcal{T})\bigr]_%
       {\mathcal{S}\in\binom{[q+t]}{q},\ \mathcal{T}\in\mathcal{E}_k},\\
  C_k(\mathcal{S},\mathcal{T})&=\det \bigl([\mathbf{h}_k, \mathbf{H}[:,\mathcal{D}_{k,\mathcal{T}}]][\mathcal{S},:]\bigr).
 \end{align}
 \label{eq:vc-decoding-system}
\end{subequations}
Thus user U$_k$ can recover all of its  missing subfiles in $\mathbf{w}^{\rm miss}_k$, if
$\mathbf{C}_k(\mathbf{H})$ has  rank $|\mathcal{E}_k|=\binom{q+t-1}{q-1}$.

For each $\mathcal{Q}=\{\mathcal Q[1]<\cdots<\mathcal Q[q-1]\}\in \binom{[q+t-1]}{q-1}$,
let $\mathcal{S}(\mathcal{Q})=\{1\}\cup\{a+1:a\in \mathcal{Q}\}$. Retaining these $\binom{q+t-1}{q-1}$
rows of $\mathbf{C}_k(\mathbf{H})$ gives the square matrix $\bigl[C_k(\mathcal{S}(\mathcal{Q}),\mathcal{T})\bigr]_%
          {\mathcal{Q}\in \binom{[q+t-1]}{q-1},\ \mathcal{T}\in\mathcal{E}_k}$, and we define its determinant as
\begin{equation}
 P_k(\mathbf{H})=\det \left(\bigl[C_k(\mathcal{S}(\mathcal{Q}),\mathcal{T})\bigr]_%
          {\mathcal{Q}\in \binom{[q+t-1]}{q-1},\ \mathcal{T}\in\mathcal{E}_k} \right).
 \label{eq:vc-selected-minor}
\end{equation}
Each entry of $\bigl[C_k(\mathcal{S}(\mathcal{Q}),\mathcal{T})\bigr]_%
{\mathcal{Q}\in \binom{[q+t-1]}{q-1},\ \mathcal{T}\in\mathcal{E}_k}$ is a $q\times q$ determinant, and therefore $P_k$ is a polynomial in the $(q+t)K$ entries of $\mathbf{H}$, with total degree at most $q\binom{q+t-1}{q-1}$. For further notational simplicity, we define 
\begin{align}
\mathbf{G}_k(\mathbf{H})= \bigl[C_k(\mathcal{S}(\mathcal{Q}),\mathcal{T})\bigr]_%
          {\mathcal{Q}\in \binom{[q+t-1]}{q-1},\ \mathcal{T}\in\mathcal{E}_k}.
\end{align}
Thus $P_k(\mathbf{H})= \det (\mathbf{G}_k(\mathbf{H})) $. To apply the Schwartz--Zippel lemma, we need to show the following lemma.
\begin{lemma}
\label{lem:vc-nonvanishing}
For every $k\in[K]$, the polynomial $P_k(\mathbf{H})$ defined in \eqref{eq:vc-selected-minor} is a non-zero polynomial.

\hfill $\square$
\end{lemma}
The proof of Lemma~\ref{lem:vc-nonvanishing} is given below by constructing a realization of $\mathbf{H}$ on which $P_k(\mathbf{H})$ exists and is nonzero. Accepting this lemma for the moment, the independently and uniformly sampled entries of $\mathbf{H}$ in Section~V-A satisfy, by the Schwartz--Zippel lemma~\cite{Schwartz1980, Zippel1979},
\begin{equation}
 \Pr\left[ \mathbf{C}_k(\mathbf{H}) \ {\rm is \ not \  full\  rank} \right]\leq \Pr[P_k(\mathbf{H})=0] \leq\frac{\deg P_k(\mathbf{H})}{p} \leq\frac{q\binom{q+t-1}{q-1}}{p}.
 \label{eq:vc-single-user-bound}
\end{equation}
Applying the union bound to all users then yields
\begin{subequations}
 \begin{align}
 &\Pr\left[ \mathbf{C}_k(\mathbf{H}) \ {\rm is   \  full\  rank}   \text{ for all }k\in[K]\right]\nonumber\\ 
 \geq&1-\sum_{k=1}^{K}
       \Pr\left[ \mathbf{C}_k(\mathbf{H}) \ {\rm is \ not \  full\  rank} \right]\\
\geq &1-\frac{Kq\binom{q+t-1}{q-1}}{p}
 \xrightarrow{p\to\infty}1.
 \end{align}
 \label{eq:vc-simultaneous-bound}
\end{subequations}
This proves Lemma~\ref{lem: SZ proof} once Lemma~\ref{lem:vc-nonvanishing} is established. 
\subsection{Proof of Lemma~\ref{lem:vc-nonvanishing} by a specific realization}
It suffices to consider user U$_1$, by the cyclic symmetry of
the placement. Indeed, for any $k\in[K]$, the simultaneous
relabeling of users and caches by
$\sigma_k(k')=\langle k+k'-1\rangle_K$ maps user~U$_1$ to user
U$_k$ and maps $\mathcal{E}_1$ bijectively to $\mathcal{E}_k$.

So in the rest of this proof, we consider user~ U$_1$. We construct a realization of the 
matrix $\mathbf{H}$ 
for which $P_1(\mathbf{H})$ exists and $P_1(\mathbf{H})\ne0$.  

User~U$_1$ can access caches C$_1$, $\ldots$, C$_L$ and needs the subfiles
$\mathbf{w}_{d_1,\mathcal{T}}$ whose supports belong to
\begin{equation}
 \mathcal{E}_1=\{\mathcal{T}\in\mathfrak{T}:\mathcal{T}\cap[L]=\emptyset\}.
 \label{eq:vc-user-one-missing-supports}
\end{equation}
We retain the cache support $\mathcal{T}$ itself as the subfile and also as the
column index of $\mathbf{C}_1(\mathbf{H})$ throughout the proof. 

Consider one set $\mathcal{T}\in\mathcal{E}_1$. Each cache node C$_{\lambda}$ with $\lambda\in \mathcal{T}$
satisfies $\lambda\geq L+1$ by
\eqref{eq:vc-user-one-missing-supports}. Hence, the set of  users who 
can obtain $\mathbf{w}_{d_1,\mathcal{T}}$ through cache node C$_{\lambda}$ is $\{\text{U}_{k'}:k'\in[\lambda-L+1:\lambda]\} \subseteq\{\text{U}_2,\ldots,\text{U}_K\}$.

The cyclic
$L$-separation of $\mathcal{T}$ makes them disjoint, so they account
for $tL$ users. Among users U$_2,\ldots,$ U$_K$, the  
$q-1$ users who do not know $\mathbf{w}_{d_1,\mathcal{T}}$ have indices in the set $\mathcal{D}_{1,\mathcal{T}}
 =\{2,\ldots,K\}\setminus\bigcup_{\lambda\in \mathcal{T}}[\lambda-L+1:\lambda].$
Write these original user indices in increasing order as $\mathcal{D}_{1,\mathcal{T}}=\{\mathcal{D}_{1,\mathcal{T}}[1],\ldots,\mathcal{D}_{1,\mathcal{T}}[q-1]\}$, where $\mathcal{D}_{1,\mathcal{T}}[i]$ denotes the $i$th smallest
 user index  in $\mathcal{D}_{1,\mathcal{T}}$.

The proof of Lemma~\ref{lem:vc-nonvanishing} contains the following steps.

$\bullet$ {\bf Step 1: A distinguished matching between the original row and column index sets.} 
To identify one term in the determinant expansion of $\mathbf{G}_1(\mathbf{H})$ (recalling that $P_1(\mathbf{H})=\det (\mathbf{G}_1(\mathbf{H}))$), we
construct a bijection $\phi:\binom{[q+t-1]}{q-1}\to\mathcal{E}_1$.
Given this bijection $\phi$, we can express $\det (\mathbf{G}_1(\mathbf{H}))$ as follows. By letting  
%To make the role of the matching explicit, let 
$\pi$ range
over all bijections $\binom{[q+t-1]}{q-1}\to\mathcal{E}_1$, and letting
$\varepsilon(\pi)\in\{1,-1\}$ be the sign of the
corresponding permutation of column positions in the fixed
row and column orders, we have 
\begin{subequations}
 \begin{align}
 \det (\mathbf{G}_1(\mathbf{H}))
 &=\sum_{\pi}\varepsilon(\pi)
      \prod_{\mathcal{Q}\in \binom{[q+t-1]}{q-1}}C_1(\mathcal{S}(\mathcal{Q}),\pi(\mathcal{Q}))\\
 &=\varepsilon(\phi)
      \prod_{\mathcal{Q}\in \binom{[q+t-1]}{q-1}}C_1(\mathcal{S}(\mathcal{Q}),\phi(\mathcal{Q})) +\sum_{\pi\ne\phi}\varepsilon(\pi)
      \prod_{\mathcal{Q}\in \binom{[q+t-1]}{q-1}}C_1(\mathcal{S}(\mathcal{Q}),\pi(\mathcal{Q})).
       \label{eq:vc-original-determinant-expansion}
 \end{align}
\end{subequations}
The product indexed by $\phi$ selects the entry in column
$\phi(\mathcal{Q})$ from each row $\mathcal{Q}$; every row and every column is
used exactly once. 
We now choose $\phi$. Under the realization constructed in Step~2, its product will be shown in Step~3 to have the unique highest degree.  For each $\mathcal{Q} \in \binom{[q+t-1]}{q-1}$, we can write its complement as $[q+t-1]\setminus \mathcal{Q}=\{\beta_1<\cdots<\beta_t\}$. We can then select the bijection $\phi$ as 
\begin{equation}
 %[q+t-1]\setminus \mathcal{Q}=\{\beta_1<\cdots<\beta_t\},
 \phi(\mathcal{Q})=\{1+\beta_j+(L-1)j:j\in[t]\}.
 \label{eq:vc-distinguished-matching}
\end{equation}
Next we will verify that $\phi$ is a valid bijection. 
\begin{itemize}
    \item[-] For any $\mathcal{Q} \in \binom{[q+t-1]}{q-1}$, we write $\lambda_j=1+\beta_j+(L-1)j$. Then it can be checked that 
\[
 \lambda_1\geq L+1,\qquad \lambda_t\leq K,\qquad
 \lambda_{j+1}-\lambda_j=\beta_{j+1}-\beta_j+L-1\geq L.
\]
The closing cyclic gap also satisfies
$K+\lambda_1-\lambda_t\geq L+1$; thus $\phi(\mathcal{Q})\in\mathcal{E}_1$.
\item[-] Conversely, for any $\mathcal{T}=\{\lambda_1<\cdots<\lambda_t\}\in\mathcal{E}_1$,
the integers $\beta_j=\lambda_j-1-(L-1)j$ where  $j\in[t]$, 
satisfy $1\leq\beta_1<\cdots<\beta_t\leq q+t-1$:
the endpoint bounds follow from $\lambda_1\geq L+1$ and $\lambda_t\leq K$,
and strict increase follows from $\lambda_{j+1}-\lambda_j\geq L$.
The complement of $\{\beta_1,\ldots,\beta_t\}$ uniquely recovers $\mathcal{Q}$. 
\end{itemize}
Thus $\phi$ is indeed a valid bijection. 

In addition, for each $\mathcal{Q} \in \binom{[q+t-1]}{q-1}$, by observing its corresponding cache-set $\phi(\mathcal{Q})=\{1+\beta_j+(L-1)j:j\in[t]\}$, we can compute the missing-user indices for
the support $ \phi(\mathcal{Q})$ (i.e., the users who are not connected to the caches in $\phi(\mathcal{Q})$) as $\mathcal{D}_{1,\phi(\mathcal{Q})}=\{\mathcal{D}_{1,\phi(\mathcal{Q})}[1],\ldots, \mathcal{D}_{1,\phi(\mathcal{Q})}[q-1]\}$, where\footnote{\label{eq:compute D}Intuitively, with $\lambda_j=1+\beta_j+(L-1)j$, we have
$\mathcal{D}_{1,\phi(\mathcal{Q})}
=[2:K]\setminus\bigcup_{j=1}^{t}[\lambda_j-L+1:\lambda_j]$.
%For $k_i=1+L\mathcal{Q}[i]-(L-1)i$, 
The identity
$|\{j:\beta_j<\mathcal{Q}[i]\}|=\mathcal{Q}[i]-i$ implies
$\lambda_j<1+L\mathcal{Q}[i]-(L-1)i$ when $\beta_j<\mathcal{Q}[i]$, and
$\lambda_j-L+1>1+L\mathcal{Q}[i]-(L-1)i$ otherwise.
Thus the $q-1$ strictly increasing indices $(1+L\mathcal{Q}[i]-(L-1)i)\in[2:K]$
lie outside all cache-covered intervals. Since
$|\mathcal{D}_{1,\phi(\mathcal{Q})}|=q-1$, $\{1+L\mathcal{Q}[i]-(L-1)i: i\in [q-1] \}$ constitute
exactly the missing-user set.}
 \begin{align}
 \mathcal{D}_{1,\phi(\mathcal{Q})}[i]
 %=2+(i-1)+L(\mathcal{Q}[i]-i)
 =1+L\mathcal{Q}[i]-(L-1)i,\qquad i\in[q-1].
 \label{eq:vc-matched-user-indices}
 \end{align}

$\bullet$ {\bf Step 2: An explicit one-parameter realization.} Let $x$ initially be an indeterminate over $\mathbb{F}_p$. At the end of the proof, we will specialize $x$ to a suitable nonzero element. 
Choose the following
$(q+t)\times K$ matrix over $\mathbb{F}_p[x]$, whose
columns correspond to the  user indices
$1,2,3,\ldots,K$:
\begin{equation}
 \mathbf{H}=
 \left[
 \begin{array}{c|cccc}
  1 & 0 & 0 & \cdots & 0\\ \hline
  0 & 1 & 1 & \cdots & 1\\
  0 & x^2 & x^3 & \cdots & x^K\\
  \vdots & \vdots & \vdots & \ddots & \vdots\\
  0 & x^{2(q+t-2)} & x^{3(q+t-2)} & \cdots & x^{K(q+t-2)}
 \end{array}
 \right].
 \label{eq:vc-explicit-realization}
\end{equation}
Its first column is $\mathbf{e}_1$, and the lower-right block has
$q+t-1$ rows. More precisely, its entries are
\begin{subequations}
 \begin{align}
 &\mathbf{h}_1=\mathbf{e}_1,\\
& h_{1,k'}=0,\qquad k'\in\{2,\ldots,K\},\\
& h_{a+1,k'}=x^{(a-1)k'},
       \qquad a\in[q+t-1],\ k'\in\{2,\ldots,K\}.
 \end{align}
 \label{eq:vc-realization-entries}
\end{subequations}
Note that the first column of $\mathbf{H}$ is $\mathbf{h}_1=\mathbf{e}_1$, with a length of $q+t$. 

We can then express $\mathbf{G}_1(\mathbf{H})$ as $\mathbf{M}(x)$, which is a polynomial of $x$. For each $\mathcal{Q}\in  \binom{[q+t-1]}{q-1}$ and each $\mathcal{T}\in\mathcal{E}_1$, $ \mathbf{M}(x)_{\mathcal{Q},\mathcal{T}}=C_1(\mathcal{S}(\mathcal{Q}),\mathcal{T})$ represents one element in the matrix $ \mathbf{M}(x)$; by
 expanding the coefficient determinant along $\mathbf{h}_1=\mathbf{e}_1$ gives
 \begin{align}
 \mathbf{M}(x)_{\mathcal{Q},\mathcal{T}}
 =C_1(\mathcal{S}(\mathcal{Q}),\mathcal{T})=\det\left( \left[
 x^{(\mathcal{Q}[i]-1)\mathcal{D}_{1,\mathcal{T}}[j]}
 \right]_{i \in [q-1],j \in [q-1]} \right). \label{eq:vc-realized-minor-matrix}
 \end{align}
 Next we need to show  that $ P_1(\mathbf{H})=\det (\mathbf{M}(x))$ is non-zero for some $x$.

$\bullet$ {\bf Step 3: A unique highest-degree term in the original ordering.} 
We will prove that $\det(\mathbf{M}(x))$ has a unique highest-degree
term whose coefficient is $1$ or $-1$.

 For each $\mathcal{Q}\in  \binom{[q+t-1]}{q-1}$ and each $\mathcal{T}\in\mathcal{E}_1$, we can expand  $\mathbf{M}(x)_{\mathcal{Q},\mathcal{T}}$ as follows: 
\begin{equation}
 \mathbf{M}(x)_{\mathcal{Q},\mathcal{T}}
 =\sum_{\sigma\in\mathfrak{S}_{q-1}}
   \operatorname{sgn}(\sigma)
   x^{\sum_{i=1}^{q-1}(\mathcal{Q}[i]-1)\mathcal{D}_{1,\mathcal{T}}[\sigma(i)]},
 \label{eq:vc-entry-expansion}
\end{equation}
where $\mathfrak{S}_{q-1}$ is the permutation group on $[q-1]$.
Note that both sequences $\mathcal{Q}[i]-1$ and $\mathcal{D}_{1,\mathcal{T}}[i]$ are
strictly increasing with $i$.
Suppose that $\sigma$ contains an
inversion, meaning that $i<j$ but $\sigma(i)>\sigma(j)$.
Then the smaller factor $\mathcal{Q}[i]-1$ is paired with the larger
user index $\mathcal{D}_{1,\mathcal{T}}[\sigma(i)]$, while the larger
factor $\mathcal{Q}[j]-1$ is paired with the smaller user index
$\mathcal{D}_{1,\mathcal{T}}[\sigma(j)]$. Exchange only the two permutation values $\sigma(i)$ and
$\sigma(j)$, defining
\[ \widetilde\sigma(\ell)=
 \begin{cases}
  \sigma(j),&\ell=i,\\
  \sigma(i),&\ell=j,\\
  \sigma(\ell),&\ell\notin\{i,j\}.
 \end{cases}\] 
All other contributions to the exponent $e(\sigma)=\sum_{i=1}^{q-1}(\mathcal Q[i]-1)\mathcal D_{1,\mathcal T}[\sigma(i)]$ remain unchanged.
Subtracting the contributions before the exchange from
those after the exchange gives
\begin{align*}
 e(\widetilde\sigma)-e(\sigma)
 &=(\mathcal{Q}[i]-1)
   \bigl(\mathcal{D}_{1,\mathcal{T}}[\sigma(j)]
        -\mathcal{D}_{1,\mathcal{T}}[\sigma(i)]\bigr)  +(\mathcal{Q}[j]-1)
   \bigl(\mathcal{D}_{1,\mathcal{T}}[\sigma(i)]
        -\mathcal{D}_{1,\mathcal{T}}[\sigma(j)]\bigr)\notag\\
 &=(\mathcal{Q}[j]-\mathcal{Q}[i])
   \bigl(\mathcal{D}_{1,\mathcal{T}}[\sigma(i)]
        -\mathcal{D}_{1,\mathcal{T}}[\sigma(j)]\bigr)>0.
\end{align*}
Thus, replacing an inverted pairing by an order-preserving pairing strictly increases the integer exponent.
Every non-identity permutation has an inversion and hence admits such a strict improvement. A maximizing permutation must therefore have no inversion; the only such permutation of $[q-1]$ is the identity. %This also covers $q-1=1$, when the identity is the only permutation. 
Consequently, the identity uniquely attains the largest exponent in~\eqref{eq:vc-entry-expansion}. Its sign is $+1$, and no other term has the same degree; thus $\mathbf{M}(x)_{\mathcal{Q},\mathcal{T}}$
has the highest degree
\begin{equation}
 \sum_{i=1}^{q-1}(\mathcal{Q}[i]-1)\mathcal{D}_{1,\mathcal{T}}[i],
 \label{eq:vc-entry-leading-degree}
\end{equation}
with coefficient $1$ multiplying this highest-degree term. Under the realization in Step~2,
\eqref{eq:vc-original-determinant-expansion} becomes
\begin{equation}
 \det (\mathbf{M}(x))
 =\sum_{\pi}\varepsilon(\pi)
       \prod_{\mathcal{Q}\in \binom{[q+t-1]}{q-1}}\mathbf{M}(x)_{\mathcal{Q},\pi(\mathcal{Q})}.
 \label{eq:vc-outer-expansion}
\end{equation}
 Following~\eqref{eq:vc-entry-leading-degree}, the product $\varepsilon(\pi)\prod_{\mathcal{Q}\in \binom{[q+t-1]}{q-1}}\mathbf{M}(x)_{\mathcal{Q},\pi(\mathcal{Q})}$ associated with $\pi$ in~\eqref{eq:vc-outer-expansion} has the highest degree 
 \begin{align}
      \sum_{\mathcal{Q}\in\binom{[q+t-1]}{q-1}} \sum_{i=1}^{q-1}(\mathcal{Q}[i]-1)\mathcal{D}_{1,\pi(\mathcal{Q})}[i],
      \label{eq:highest degree of pi}
      %E(\mathcal{Q},\pi(\mathcal{Q}))
 \end{align}
 with coefficient $\varepsilon(\pi)$ multiplied this highest-degree term.

For the selected   matching $\phi$ in~\eqref{eq:vc-distinguished-matching}, the highest degree of the product $\prod_{\mathcal{Q}\in \binom{[q+t-1]}{q-1}}\mathbf{M}(x)_{\mathcal{Q},\phi(\mathcal{Q})}$  is
 \begin{align}
 \delta:= %=\Delta(\phi)
 \sum_{\mathcal{Q}\in\binom{[q+t-1]}{q-1}}\sum_{i=1}^{q-1}
       (\mathcal{Q}[i]-1)\mathcal{D}_{1,\phi(\mathcal{Q})}[i] =\sum_{\mathcal{Q}\in\binom{[q+t-1]}{q-1}}\sum_{i=1}^{q-1}
       (\mathcal{Q}[i]-1)\bigl(1+L\mathcal{Q}[i]-(L-1)i\bigr).
       \label{eq:vc-witness-degree}
 \end{align}
For each $i\in [q-1]$ and each $r\in\{1,2\}$, both $\phi$ and $\pi$
visit each support in $\mathcal{E}_1$ exactly once, so
\begin{equation}
 \sum_{\mathcal{Q}\in \binom{[q+t-1]}{q-1}}\bigl(\mathcal{D}_{1,\phi(\mathcal{Q})}[i]\bigr)^{r}
 =\sum_{\mathcal{T}\in\mathcal{E}_1}\bigl(\mathcal{D}_{1,\mathcal{T}}[i]\bigr)^{r}
 =\sum_{\mathcal{Q}\in \binom{[q+t-1]}{q-1}}\bigl(\mathcal{D}_{1,\pi(\mathcal{Q})}[i]\bigr)^{r}.
 \label{eq:vc-matching-moments}
\end{equation}
Also, \eqref{eq:vc-matched-user-indices} gives
\begin{equation}
 \mathcal{Q}[i]-1=
 \frac{\mathcal{D}_{1,\phi(\mathcal{Q})}[i]+(L-1)i-L-1}{L}.
 \label{eq:vc-row-exponent-identity}
\end{equation}
Then using \eqref{eq:vc-matching-moments}, we compute the difference between the degrees in~\eqref{eq:vc-witness-degree} and~\eqref{eq:highest degree of pi}  as follows:
\begin{subequations}
\label{eq:vc-unique-maximizer}
\begin{align}
 &\delta-\sum_{\mathcal{Q}\in\binom{[q+t-1]}{q-1}} \sum_{i=1}^{q-1}(\mathcal{Q}[i]-1)\mathcal{D}_{1,\pi(\mathcal{Q})}[i]\nonumber\\
 =&\sum_{\mathcal{Q}\in\binom{[q+t-1]}{q-1}}\sum_{i=1}^{q-1}
 (\mathcal{Q}[i]-1)
 \bigl(\mathcal{D}_{1,\phi(\mathcal{Q})}[i]
       -\mathcal{D}_{1,\pi(\mathcal{Q})}[i]\bigr)
 \label{eq:vc-unique-maximizer-difference}\\
= &\frac1L\sum_{\mathcal{Q}\in\binom{[q+t-1]}{q-1}}\sum_{i=1}^{q-1}
 \left[
  \bigl(\mathcal{D}_{1,\phi(\mathcal{Q})}[i]\bigr)^2
  -\mathcal{D}_{1,\phi(\mathcal{Q})}[i]\mathcal{D}_{1,\pi(\mathcal{Q})}[i]
 \right]
 \label{eq:vc-unique-maximizer-first-moment}\\
= &\frac1{2L}\sum_{\mathcal{Q}\in\binom{[q+t-1]}{q-1}}\sum_{i=1}^{q-1}
 \bigl(\mathcal{D}_{1,\phi(\mathcal{Q})}[i]
       -\mathcal{D}_{1,\pi(\mathcal{Q})}[i]\bigr)^2,
 \label{eq:vc-unique-maximizer-square}
\end{align}
\end{subequations}
where \eqref{eq:vc-unique-maximizer-first-moment} follows by
substituting \eqref{eq:vc-row-exponent-identity} and using
\eqref{eq:vc-matching-moments} with $r=1$, which gives
\[\sum_{\mathcal{Q}\in\binom{[q+t-1]}{q-1}}
(\mathcal{D}_{1,\phi(\mathcal{Q})}[i]-\mathcal{D}_{1,\pi(\mathcal{Q})}[i])=0\]
for each fixed $i$ and thus cancels the terms multiplied
by $(L-1)i-L-1$; and
\eqref{eq:vc-unique-maximizer-square} follows from
\eqref{eq:vc-matching-moments} with $r=2$, which equates
the two sums of squares.

The sum of squares in~\eqref{eq:vc-unique-maximizer-square} vanishes only if $\mathcal{D}_{1,\phi(\mathcal{Q})}=\mathcal{D}_{1,\pi(\mathcal{Q})}$ for every $\mathcal{Q}$.
By Step~1, distinct supports have distinct missing-user
lists, so this happens only when $\pi=\phi$.
Thus every other product has degree strictly below $\delta$.
 Hence, the term with the highest degree of $x$ in $\det (\mathbf{M}(x))$ is with the degree equal to $\delta$, corresponding to the bijection $\phi$; mathematically, we have 
\begin{equation}
 \det (\mathbf{M}(x))=\varepsilon(\phi)x^\delta
                 +\text{terms of degree less than }\delta,
 \qquad \varepsilon(\phi)\in\{1,-1\}.
 \label{eq:vc-leading-determinant}
\end{equation}
Since the value of $\delta$ is finite and $x\in \mathbb{F}_{p}$ with $p \to \infty$, we can easily select $x$ such that $\det (\mathbf{M}(x)) \neq 0$.

As a result, we find one realization of $\mathbf{H}$ where $P_1(\mathbf{H})$ exists and is not equal to zero; thus we can prove Lemma~\ref{lem:vc-nonvanishing}. 

\bibliographystyle{IEEEtran}
\bibliography{reference}

@ARTICLE{WCWL,
  author={Wang, Jinyu and Cheng, Minquan and Wu, Youlong and Li, Xianxian},
  journal={IEEE Transactions on Communications}, 
  title={Multi-Access Coded Caching With Optimal Rate and Linear Subpacketization Under PDA and Consecutive Cyclic Placement}, 
  year={2023},
  volume={71},
  number={6},
  pages={3178-3190},
  doi={10.1109/TCOMM.2023.3256417}
}

@INPROCEEDINGS{LCWC,
  author={Li, Mengyuan and Cheng, Minquan and Wan, Kai and Caire, Giuseppe},
  booktitle={2026 IEEE International Symposium on Information Theory (ISIT)}, 
  title={A New Construction Structure on Multi-access Coded Caching with Linear Subpacketization: Cyclic Multi-Access Non-Half-Sum Disjoint Packing}, 
  year={2026},
  volume={},
  number={},
  pages={1-6},
  doi={10.1109/ISIT62367.2026.11653665}}

@article{MN,
  author        = "Maddah-Ali, Mohammad Ali and Niesen, Urs",
  title         = "Fundamental Limits of Caching",
  journal       = "IEEE Trans. Inf. Theory",
  volume        = "60",
  number        = "5",
  month         = "May",
  year          = "2014",
  pages         = "2856-2867"
}

@article{Ghasemi2017,
  author  = {Ghasemi, Hooshang and Ramamoorthy, Aditya},
  title   = {Improved Lower Bounds for Coded Caching},
  journal = {IEEE Transactions on Information Theory},
  year    = {2017},
  volume  = {63},
  number  = {7},
  pages   = {4388--4413},
  doi     = {10.1109/TIT.2017.2705166}
}

@article{Yu2019Converse,
  author  = {Yu, Qian and Maddah-Ali, Mohammad Ali and Avestimehr, A. Salman},
  title   = {Characterizing the Rate-Memory Tradeoff in Cache Networks Within a Factor of 2},
  journal = {IEEE Transactions on Information Theory},
  volume  = {65},
  number  = {1},
  pages   = {647--663},
  year    = {2019},
  doi     = {10.1109/TIT.2018.2870566}
}

@ARTICLE{Wang2018Converse,
  author={Wang, Chien-Yi and Saeedi Bidokhti, Shirin and Wigger, Mich{\`e}le},
  journal={IEEE Transactions on Information Theory}, 
  title={Improved Converses and Gap Results for Coded Caching}, 
  year={2018},
  volume={64},
  number={11},
  pages={7051-7062},
  doi={10.1109/TIT.2018.2856885}}

@article{Tian2018,
  author  = {Tian, Chao},
  title   = {Symmetry, Outer Bounds, and Code Constructions: A Computer-Aided Investigation on the Fundamental Limits of Caching},
  journal = {Entropy},
  volume  = {20},
  number  = {8},
  pages   = {603},
  year    = {2018},
  doi     = {10.3390/e20080603}
}

@INPROCEEDINGS{WTP2016,
  author        = "Kai Wan and Tuninetti, Daniela and Piantanida, Pablo",
  title         = "On the optimality of uncoded cache placement",
  booktitle     = "Proc. IEEE Inf. Theory Workshop (ITW)",
  month         = "Sep.",
  year          = "2016",
  pages         = "161-165"
  
}

@article{WTP2020,
  author        = "Wan, Kai and Tuninetti, Daniela and Piantanida, Pablo",
  title         = "An index coding approach to caching with uncoded cache placement",
  journal       = "IEEE Trans. Inf. Theory",
  volume        = "66",
  number        = "3",
  month         = "Mar.",
  year          = "2020",
  pages         = "1318-1332"
}

@ARTICLE{ZYST,
  author={Bar-Yossef, Ziv and Birk, Yitzhak and Jayram, T. S. and Kol, Tomer},
  journal={IEEE Transactions on Information Theory}, 
  title={Index Coding With Side Information}, 
  year={2011},
  volume={57},
  number={3},
  pages={1479-1494},
  doi={10.1109/TIT.2010.2103753}}

@article{YMA2018,
  author        = "Yu, Qian and Maddah-Ali, Mohammad Ali and Avestimehr, A. Salman",
  title         = "The exact rate-memory tradeoff for caching with uncoded prefetching",
  journal       = "IEEE Trans. Inf. Theory",
  volume        = "64",
  number        = "2",
  month         = "Feb.",
  year          = "2018",
  pages         = "1281-1296"
}

@inproceedings{zippel1979,
  title={Probabilistic algorithms for sparse polynomials},
  author={Zippel, Richard},
  booktitle={International symposium on symbolic and algebraic manipulation},
  pages={216--226},
  year={1979},
  organization={Springer}
}

@article{Schwartz1980,
  title={Fast probabilistic algorithms for verification of polynomial identities},
  author={Schwartz, Jacob T},
  journal={Journal of the ACM (JACM)},
  volume={27},
  number={4},
  pages={701--717},
  year={1980},
  publisher={ACM New York, NY, USA}
}

@article{SJTLD,
  author        = "Shanmugam, Karthikeyan and Ji, Mingyue and Tulino, Antonia M. and Llorca, Jaime and Dimakis, Alexandros G.",
  title         = "Finite-length analysis of caching-aided coded multicasting",
  journal       = "IEEE Trans. Inf. Theory",
  volume        = "62",
  number        = "10",
  month         = "Oct.",
  year          = "2016",
  pages         = "5524-5537"
}

@article{YCTC,
  author        = "Yan, Qifa and Cheng, Minquan and Tang, Xiaohu and Chen, Qingchun",
  title         = "On the placement delivery array design for centralized coded caching scheme",
  journal       = "IEEE Trans. Inf. Theory",
  volume        = "63",
  number        = "9",
  month         = "Sep.",
  year          = "2017",
  pages         = "5821-5833"  
}

@article{SZG,
  author        = "Shangguan, Chong and Zhang, Yiwei and Ge, Gennian",
  title         = "Centralized coded caching schemes: A hypergraph theoretical approach",
  journal       = "IEEE Trans. Inf. Theory",
  volume        = "64",
  number        = "8",
  month         = "Aug.",
  year          = "2018",
  pages         = "5755-5766"  
}

@article{CJTY,
  author        = "Cheng, Minquan and Jiang, Jing and Tang, Xiaohu and Yan, Qifa",
  title         = "Some variant of known coded caching schemes with good performance",
  journal       = "IEEE Trans. Commun.",
  volume        = "68",
  number        = "3",
  month         = "Mar.",
  year          = "2020",
  pages         = "1370-1377"  
}

@article{YTCC,
  author        = "Yan, Qifa and Tang, Xiaohu and Chen, Qingchun and Cheng, Minquan",
  title         = "Placement delivery array design through strong edge coloring of bipartite graphs",
  journal       = "IEEE Commun. Lett.",
  volume        = "22",
  number        = "2",
  month         = "Feb.",
  year          = "2018",
  pages         = "236-239"  
}

@inproceedings{K,
  author        = "Krishnan, Prasad",
  title         = "Coded caching via line graphs of bipartite graphs",
  booktitle     = "Proc. IEEE Inf. Theory Workshop (ITW)",
  month         = "Nov.",
  year          = "2018",
  pages         = "1-5"
}

@inproceedings{ASK,
  author        = "Agrawal, Shailja and Sushena Sree, K V and Krishnan, Prasad",
  title         = "Coded caching based on combinatorial designs",
  booktitle     = "Proc. IEEE Int. Symp. Inf. Theory (ISIT)",
  month         = "Jul.",
  year          = "2019",
  pages         = "1227-1231"
}

@article{CJWY,
  author        = "Cheng, Minquan and Jiang, Jing and Wang, Qiang and Yao, Youzhi",
  title         = "A generalized grouping scheme in coded caching",
  journal       = "IEEE Trans. Commun.",
  volume        = "67",
  number        = "5",
  month         = "May",
  year          = "2019",
  pages         = "3422-3430"  
}

@article{CJYT,
  author        = "Cheng, Minquan and Jiang, Jing and Yan, Qifa and Tang, Xiaohu",
  title         = "Constructions of Coded Caching Schemes With Flexible Memory Size",
  journal       = "IEEE Trans. Commun.",
  volume        = "67",
  number        = "6",
  month         = "Jun.",
  year          = "2019",
  pages         = "4166-4176"  
}

@article{CWZW,
  author={Cheng, Minquan and Wang, Jinyu and Zhong, Xi and Wang, Qiang},
  journal={IEEE Trans. Inf. Theory}, 
  title={A Framework of Constructing Placement Delivery Arrays for Centralized Coded Caching}, 
  month={Nov.},
  year={2021},
  volume={67},
  number={11},
  pages={7121-7131},
  doi={10.1109/TIT.2021.3112492}}

@book{L,
  title={Introduction to Coding Theory},
  author={J. H. van Lint},
  publisher={Springer-Verlag},
  year={1999}
}

@InProceedings{R,
author="Blackburn, Simon R.
and Burmester, Mike
and Desmedt, Yvo
and Wild, Peter R.",
editor="Maurer, Ueli",
title="Efficient Multiplicative Sharing Schemes",
booktitle="Advances in Cryptology --- EUROCRYPT '96",
year="1996",
publisher="Springer Berlin Heidelberg",
address="Berlin, Heidelberg",
pages="107--118",
}

@ARTICLE{HKD,
  author={Hachem, Jad and Karamchandani, Nikhil and Diggavi, Suhas N.},
  journal={IEEE Trans. Inf. Theory}, 
  title={Coded Caching for Multi-level Popularity and Access}, 
  month={May},
  year={2017},
  volume={63},
  number={5},
  pages={3108-3141},
  doi={10.1109/TIT.2017.2664817}}

@ARTICLE{RK,
  author={Reddy, Kota Srinivas and Karamchandani, Nikhil},
  journal={IEEE Trans. Commun.}, 
  title={Rate-Memory Trade-off for Multi-Access Coded Caching With Uncoded Placement}, 
  month={Jun.},
  year={2020},
  volume={68},
  number={6},
  pages={3261-3274},
  doi={10.1109/TCOMM.2020.2980817}}

@INPROCEEDINGS{MR,
  author={Mahesh, Anjana Ambika and Sundar Rajan, B.}, 
  booktitle={Proc. IEEE Inf. Theory Workshop (ITW)}, 
  title={A Coded Caching Scheme with Linear Sub-packetization and its Application to Multi-Access Coded Caching}, 
  month={Apr.},
  year={2021},
  volume={},
  number={},
  pages={1-5},
  doi={10.1109/ITW46852.2021.9457641}}

@ARTICLE{SR,
  author={Sasi, Shanuja and Rajan, B. Sundar},
  journal={IEEE Trans. Commun.}, 
  title={Multi-Access Coded Caching Scheme With Linear Sub-Packetization Using {PDA}s}, 
  month={Dec.},
  year={2021},
  volume={69},
  number={12},
  pages={7974-7985},
  doi={10.1109/TCOMM.2021.3116229}}

@INPROCEEDINGS{Das2022CRD,
  author={Das, Niladri and Rajan, B. Sundar},
  booktitle={2022 IEEE International Symposium on Information Theory (ISIT)},
  title={Multi-Access Coded Caching Schemes from Maximal Cross Resolvable Designs},
  year={2022},
  pages={1719-1724},
  doi={10.1109/ISIT50566.2022.9834353}
}

@ARTICLE{CWLZC,
  author={Cheng, Minquan and Wan, Kai and Liang, Dequan and Zhang, Mingming and Caire, Giuseppe},
  journal={IEEE Trans. Commun.}, 
  title={A Novel Transformation Approach of Shared-Link Coded Caching Schemes for Multiaccess Networks}, 
  month={Nov.},
  year={2021},
  volume={69},
  number={11},
  pages={7376-7389},
  doi={10.1109/TCOMM.2021.3104035}}

@ARTICLE{RK2,
  author={Reddy, Kota Srinivas and Karamchandani, Nikhil},
  journal={IEEE J. Sel. Areas Commun.}, 
  title={Structured Index Coding Problem and Multi-Access Coded Caching}, 
  month={Dec.},
  year={2021},
  volume={2},
  number={4},
  pages={1266-1281},
  doi={10.1109/JSAIT.2021.3126663}}

@INPROCEEDINGS{Sengupta2015,
  author={Sengupta, Avik and Tandon, Ravi and Clancy, T. Charles},
  booktitle={2015 IEEE International Symposium on Information Theory (ISIT)}, 
  title={Improved approximation of storage-rate tradeoff for caching via new outer bounds}, 
  year={2015},
  volume={},
  number={},
  pages={1691-1695},
  doi={10.1109/ISIT.2015.7282744}}

@INPROCEEDINGS{SPE,
  author={Serbetci, Berksan and Parrinello, Emanuele and Elia, Petros},
  booktitle={Proc. IEEE Inf. Theory Workshop (ITW)}, 
  title={Multi-access coded caching: gains beyond cache-redundancy}, 
  month={Aug.},
  year={2019},
  volume={},
  number={},
  pages={1-5},
  doi={10.1109/ITW44776.2019.8989128}}

@article{SR_arX2021,
  author   = {Shanuja Sasi and B. Sundar Rajan},
  title    = {An Improved Multi-access Coded Caching with Uncoded Placement},
  journal  = {arXiv:2009.05377v3},
  month= {Feb.},
  year    = {2021}
}

@ARTICLE{Katyal2021CRD,
  author={Katyal, Digvijay and Muralidhar, Pooja Nayak and Rajan, B. Sundar},
  journal={IEEE Transactions on Communications},
  title={Multi-Access Coded Caching Schemes From Cross Resolvable Designs},
  year={2021},
  volume={69},
  number={5},
  pages={2997-3010},
  doi={10.1109/TCOMM.2021.3053048}
}

@ARTICLE{NR,
  author={Namboodiri, K. K. Krishnan and Rajan, B. Sundar},
  journal={IEEE Trans. Commun.}, 
  title={Improved Lower Bounds for Multi-Access Coded Caching}, 
  month={Jul.},
  year={2022},
  volume={70},
  number={7},
  pages={4454-4468},
  doi={10.1109/TCOMM.2022.3174890}}

@INPROCEEDINGS{Vijith2023MACC,
  author={{Vijith Kumar K. P.} and Rai, Brijesh Kumar and Jacob, Tony},
  booktitle={2023 IEEE Information Theory Workshop (ITW)},
  title={The Optimal Rate Memory Tradeoff in Multi-Access Coded Caching: Large Cache Size},
  year={2023},
  pages={165-169},
  doi={10.1109/ITW55543.2023.10161659}
}

@ARTICLE{Cheng2025Topology,
  author={Cheng, Minquan and Wan, Kai and Elia, Petros and Caire, Giuseppe},
  journal={IEEE Transactions on Information Theory},
  title={Coded Caching Schemes for Multiaccess Topologies via Combinatorial Design},
  year={2025},
  volume={71},
  number={7},
  pages={5027-5048},
  doi={10.1109/TIT.2025.3567008}
}

@ARTICLE{Zhang2024TwoDim,
  author={Zhang, Mingming and Wan, Kai and Cheng, Minquan and Caire, Giuseppe},
  journal={IEEE Transactions on Information Theory},
  title={Coded Caching for Two-Dimensional Multi-Access Networks With Cyclic Wrap Around},
  year={2024},
  volume={70},
  number={1},
  pages={30-50},
  doi={10.1109/TIT.2023.3332597}
}

@ARTICLE{Brunero2023Combinatorial,
  author={Brunero, Federico and Elia, Petros},
  journal={IEEE Transactions on Information Theory},
  title={Fundamental Limits of Combinatorial Multi-Access Caching},
  year={2023},
  volume={69},
  number={2},
  pages={1037-1056},
  doi={10.1109/TIT.2022.3193723}
}

@book{CoverThomas2006,
  author    = {Thomas M. Cover and Joy A. Thomas},
  title     = {Elements of Information Theory},
  edition   = {2},
  publisher = {Wiley-Interscience},
  address   = {Hoboken, NJ, USA},
  year      = {2006}
}

@INPROCEEDINGS{Peter2023MultiAntenna,
  author={Peter, Elizabath and Rajan, B. Sundar},
  booktitle={2023 IEEE International Symposium on Information Theory (ISIT)},
  title={Multi-antenna Coded Caching with Combinatorial Multi-access Networks},
  year={2023},
  pages={126-131},
  doi={10.1109/ISIT54713.2023.10206943}
}

@ARTICLE{Cheng2024MAMISO,
  author={Cheng, Minquan and Bai, Wenli and Wu, Xianzhang and Chen, Shuwu},
  journal={IEEE Wireless Communications Letters},
  title={Design of Coded Caching Scheme for MAMISO Networks With Improved DoF},
  year={2024},
  volume={13},
  number={5},
  pages={1478-1482},
  doi={10.1109/LWC.2024.3376401}
}

@ARTICLE{Peter2025MultiAntenna,
  author={Peter, Elizabath and Namboodiri, K. K. Krishnan and Rajan, B. Sundar},
  journal={IEEE Transactions on Communications},
  title={Multi-Antenna Coded Caching for Multi-Access Networks With Cyclic Wrap-Around},
  year={2025},
  volume={73},
  number={8},
  pages={6423-6439},
  doi={10.1109/TCOMM.2025.3534461}
}

@INPROCEEDINGS{KK2023,
  author={Reddy, Kota Srinivas and Karamchandani, Nikhil},
  booktitle={2023 IEEE International Symposium on Information Theory (ISIT)},
  title={Multi-access Coded Caching with Linear Subpacketization},
  year={2023},
  pages={418-423},
  doi={10.1109/ISIT54713.2023.10206602}
}

@INPROCEEDINGS{WCWC2022ISIT,
  author={Wang, Jinyu and Cheng, Minquan and Wan, Kai and Caire, Giuseppe},
  booktitle={Proc. IEEE Int. Symp. Inf. Theory (ISIT)}, 
  title={A Novel Framework for Coded Caching via Cartesian Product with Reduced Subpacketization}, 
  month={Jun.},
  year={2022},
  volume={},
  number={},
  pages={1300-1305},
  doi={10.1109/ISIT50566.2022.9834638}}

@ARTICLE{CWWC2026,
  author={Cheng, Minquan and Wei, Huimei and Wan, Kai and Caire, Giuseppe},
  journal={IEEE Transactions on Information Theory}, 
  title={A New Construction Structure on Coded Caching With Linear Subpacketization: Non-Half-Sum Disjoint Packing}, 
  year={2026},
  volume={72},
  number={6},
  pages={3753-3770},
  doi={10.1109/TIT.2026.3685007}}

@INPROCEEDINGS{Ajaykrishnan2015,
  author={Ajaykrishnan N. and Prem, Navya S. and Prabhakaran, Vinod M. and Vaze, Rahul},
  booktitle={2015 Twenty First National Conference on Communications (NCC)}, 
  title={Critical database size for effective caching}, 
  year={2015},
  volume={},
  number={},
  pages={1-6},
  doi={10.1109/NCC.2015.7084871}}

@article{Krishnan2021,
  title={An umbrella converse for data exchange: Applied to caching, computing, and shuffling},
  author={Krishnan, Prasad and Natarajan, Lakshmi and Lalitha, Vadlamani},
  journal={Entropy},
  volume={23},
  number={8},
  pages={985},
  year={2021},
  publisher={MDPI}
}

@inproceedings{Ji2013,
  author    = {Mingyue Ji and Giuseppe Caire and Andreas F. Molisch},
  title     = {Fundamental Limits of Distributed Caching in {D2D} Wireless Networks},
  booktitle = {Proc. IEEE Information Theory Workshop (ITW)},
  pages     = {1--5},
  year      = {2013},
  doi       = {10.1109/ITW.2013.6691247}
}

@article{Karamchandani2016,
  author  = {Nikhil Karamchandani and Urs Niesen and Mohammad Ali Maddah-Ali and Suhas N. Diggavi},
  title   = {Hierarchical Coded Caching},
  journal = {IEEE Transactions on Information Theory},
  volume  = {62},
  number  = {6},
  pages   = {3212--3229},
  year    = {2016},
  doi     = {10.1109/TIT.2016.2557804}
}

@inproceedings{Hachem2015,
  author    = {Jad Hachem and Nikhil Karamchandani and Suhas N. Diggavi},
  title     = {Content Caching and Delivery over Heterogeneous Wireless Networks},
  booktitle = {Proc. IEEE INFOCOM},
  pages     = {756--764},
  year      = {2015},
  doi       = {10.1109/INFOCOM.2015.7218445}
}

@inproceedings{Ji2015Combination,
  author    = {Mingyue Ji and Ming Fai Wong and Antonia M. Tulino and Jaime Llorca and Giuseppe Caire and Michelle Effros and Michael Langberg},
  title     = {On the Fundamental Limits of Caching in Combination Networks},
  booktitle = {Proc. IEEE Workshop on Signal Processing Advances in Wireless Communications (SPAWC)},
  pages     = {695--699},
  year      = {2015},
  doi       = {10.1109/SPAWC.2015.7227127}
}

@article{Girgis2017,
  title={Degrees of freedom of interference networks with
transmitter-side caches},
  author={Girgis, A. M. and Ercetin, O. and Nafie, M. and ElBatt, T.},
  journal={arXiv preprint arXiv:1712.05957},
  year={2017}
}

@ARTICLE{Wanscalar,
  author={Wan, Kai and Sun, Hua and Ji, Mingyue and Tuninetti, Daniela and Caire, Giuseppe},
  journal={IEEE Transactions on Information Theory}, 
  title={On the Optimal Load-Memory Tradeoff of Cache-Aided Scalar Linear Function Retrieval}, 
  year={2021},
  volume={67},
  number={6},
  pages={4001-4018}}

@ARTICLE{Mascalar,
  author={Ma, Yinbin and Tuninetti, Daniela},
  journal={IEEE Journal on Selected Areas in Information Theory}, 
  title={A General Coded Caching Scheme for Scalar Linear Function Retrieval}, 
  year={2022},
  volume={3},
  number={2},
  pages={321-336}}

@INPROCEEDINGS{Wang2016Converse,
  author={Wang, Chien-Yi and Lim, Sung Hoon and Gastpar, Michael},
  booktitle={2016 Information Theory and Applications Workshop (ITA)}, 
  title={A new converse bound for coded caching}, 
  year={2016},
  volume={},
  number={},
  pages={1-6},
  doi={10.1109/ITA.2016.7888186}}

@ARTICLE{ZCW2021,
  author={Zhong, Xi and Cheng, Minquan and Wei, Ruizhong},
  journal={IEEE Transactions on Communications}, 
  title={Coded Caching Schemes With Linear Subpacketizations}, 
  year={2021},
  volume={69},
  number={6},
  pages={3628-3637},
  doi={10.1109/TCOMM.2021.3062667}
}

@Article{AST2022,
  author  = {Aravind, V. R. and Sarvepalli, Pradeep Kiran and Thangaraj, Andrew},
  journal = {IEEE Transactions on Communications},
  title   = {Lifting Constructions of PDAs for Coded Caching With Linear Subpacketization},
  year    = {2022},
  number  = {12},
  pages   = {7817-7829},
  volume  = {70},
  doi     = {10.1109/TCOMM.2022.3216641},
}

@ARTICLE{MW2020,
  author={Michel, Jerod and Wang, Qi},
  journal={IEEE Trans. Commun.}, 
  title={Placement Delivery Arrays From Combinations of Strong Edge Colorings}, 
  month={Oct.},
  year={2020},
  volume={68},
  number={10},
  pages={5953-5964},
  doi={10.1109/TCOMM.2020.3005425}}

@inproceedings{STD2017,
  author        = "Shanmugam, Karthikeyan and Tulino, Antonia M. and Dimakis, Alexandros G.",
  title         = "Coded caching with linear subpacketization is possible using Ruzsa-Szem\'{e}redi graphs",
  booktitle     = "Proc. IEEE Int. Symp. Inf. Theory (ISIT)",
  month         = "Jun.",
  year          = "2017",
  pages         = "1237-1241"
}

@inproceedings{K2018,
  author        = "Krishnan, Prasad",
  title         = "Coded caching via line graphs of bipartite graphs",
  booktitle     = "Proc. IEEE Inf. Theory Workshop (ITW)",
  month         = "Nov.",
  year          = "2018",
  pages         = "1-5"
}

@article{TR2018,
  author        = "Tang, Li and Ramamoorthy, Aditya",
  title         = "Coded caching schemes with reduced subpacketization from linear block codes",
  journal       = "IEEE Trans. Inf. Theory",
  volume        = "64",
  number        = "4",
  month         = "Apr.",
  year          = "2018",
  pages         = "3099-3120"  
}

@ARTICLE{CKSB2021,
  author={Chittoor, Hari Hara Suthan and Krishnan, Prasad and Sree, K. V. Sushena and Mamillapalli, Bhavana},
  journal={IEEE Trans. Inf. Theory}, 
  title={Subexponential and Linear Subpacketization Coded Caching via Projective Geometry}, 
  month={Sep.},
  year={2021},
  volume={67},
  number={9},
  pages={6193-6222},
  doi={10.1109/TIT.2021.3095471}}

@article{WCWC2025,
  author  = {Wang, Jinyu and Cheng, Minquan and Wan, Kai and Caire, Giuseppe},
  title   = {{PDA} Construction via Union of Cache Configurations From Cartesian Product},
  journal = {IEEE Transactions on Communications},
  year    = {2026},
  doi     = {10.1109/TCOMM.2026.3725096}
}

\end{document}